\documentclass{emulateapj}

\usepackage{natbib}
\usepackage{longtable}

\shorttitle{Spitzer Detected Starbursts in the Cl1604 Supercluster}
\shortauthors{Kocevski et al.}

\begin{document}

\title{Obscured Starburst Activity in High Redshift Clusters and Groups}

\author{Dale D. Kocevski, Brian C. Lemaux\altaffilmark{1}, Lori M. Lubin\altaffilmark{1}, Roy Gal\altaffilmark{2}, Elizabeth J. McGrath, Christopher D. Fassnacht\altaffilmark{1}, Gordon K. Squires\altaffilmark{3}, Jason A. Surace\altaffilmark{3}, Mark Lacy\altaffilmark{3}}

\affil{University of California Observatories/Lick Observatory, University of
  California, Santa Cruz, CA 95064}

\altaffiltext{1}{Department of Physics, University of California, Davis, 1 Shields Avenue,  Davis, CA 95616}
\altaffiltext{2}{Institute for Astronomy, University of Hawaii, 2680 Woodlawn Dr., Honolulu, HI 96822}
\altaffiltext{3}{\emph{Spitzer} Science Center, M/S 220-6, California 
                 Institute of Technology, 1200 East California Blvd, Pasadena, CA 91125}
\email{kocevski@physics.ucdavis.edu}

\begin{abstract}
Using \emph{Spitzer}-MIPS $24\mu$m imaging and extensive \emph{Keck}
spectroscopy we examine the nature of the obscured star forming population in
three clusters and three groups at $z\sim0.9$.  These six systems are the
primary components of the Cl1604 supercluster, the largest structure imaged
by \emph{Spitzer} at redshifts approaching unity.  We find that the average
density of $24\mu$m-detected galaxies within the Cl1604 clusters is nearly
twice that of the surrounding field and that this overdensity scales with the
cluster's dynamical state.  The $24\mu$m-bright members often appear
optically unremarkable and exhibit only moderate [OII] line emission due to
severe obscuration. Their spatial distribution suggests they are an infalling
population, but an examination of their spectral properties, morphologies and
optical colors indicate they are not simply analogs of the field population
that have yet to be quenched.  Using stacked composite spectra, we find the
$24\mu$m-detected cluster and group galaxies exhibit elevated levels of
Balmer absorption compared to galaxies undergoing normal, continuous star
formation.  A similar excess is not observed in field galaxies with
equivalent infrared luminosities, indicating a greater fraction of the
detected cluster and group members have experienced a burst of star
formation in the recent past compared to their counterparts in the field.       
Our results suggest that gas-rich galaxies at high redshift experience a
temporary increase in their star formation activity as they assemble into
denser environments.  Using \emph{HST}-ACS imaging we find that disturbed
morphologies are common among the 24$\mu$m-detected cluster and group members
and become more prevalent in regions of higher galaxy density.  We conclude
that mergers are the dominant triggering mechanism responsible for the
enhanced star formation found in the Cl1604 groups, while a mix of harassment
and mergers are likely driving the activity of the cluster galaxies.
\end{abstract}

\keywords{galaxies: starburst --- galaxies: clusters: general --- galaxies: evolution}

\section{Introduction}

It is well established that in the local Universe dense environments such as
galaxy clusters are generally hostile to star formation as they predominantly
contain passively evolving, early-type galaxies. 
At higher redshifts, though, there is increasing evidence that star formation
is far more ubiquitous in such environments.  
Among the earliest indications was the detection of an increased fraction of blue,
late-type galaxies in moderate-redshift clusters, otherwise known as
the Butcher-Oemler effect (Butcher \& Oemler 1978, 1984; Dressler et al.~1997).  
More recent observations have revealed this increase to be part of an overall
evolution in the morphology-density relationship such that spirals are more
common in dense environments at higher redshifts than they are locally,
largely at the expense of the S0 population (Postman et al.~2005).  
In conjunction with this morphological evolution, spectroscopic studies have
revealed a large population of post-starburst galaxies in distant
clusters that are largely absent at lower redshifts (Dressler \& Gunn 1983;
Dressler et al.~1999; Poggianti et al.~1999). 

These post-starburst systems, also known as k+a galaxies, exhibit Balmer
absorption features due to the presence of recently formed ($<$1 Gyr) A-type 
stars, but lack emission lines that are indicative of ongoing star formation, such
as the [OII] doublet at 3727\AA.  While such features can be produced by the
truncation of normal star formation, it has been shown that spectra with
strong Balmer absorption lines require a burst of 
star formation prior to a rapid quenching (Poggianti 2004).  The existence of
a large population of such systems suggest that galaxies in higher-redshift clusters
experience a period of increased star formation activity that is not observed
in their low redshift counterparts.     

Until recently, efforts to study starburst galaxies within clusters have proven difficult
because surveys in the UV and optical failed to detect
the starbursting progenitors of the post-starburst population found by  
Dressler et al.~(1999) (e.g.~Balogh et al.~1997). 
It is now well appreciated that dust can heavily obscure such systems (Silva
et al. 1998) resulting in severely underestimated star formation rates (SFR)
when relying solely on optical line diagnostics (Kennicutt 1998, Kewley et
al. 2004).  Infrared (IR) observations, on the other hand, are sensitive to
stellar radiation reprocessed by dust and therefore provide a means to
penetrate this obscuration.    

\begin{center}
\begin{deluxetable*}{cccccccc}
\tabletypesize{\scriptsize}
\tablewidth{0pt}
\tablecaption{Properties of Galaxy Clusters and Groups in the Cl1604 Supercluster \label{tab-cl_prop}}
\tablecolumns{8}
\tablehead{\colhead{} & \colhead{} & \colhead{RA}  & \colhead{Dec}  &
  \colhead{}  & \colhead{$\sigma_{\rm v}$} & \colhead{$R_{\rm vir}$} & \colhead{$N_{\rm gal}$} \\
\colhead{ID} & \colhead{Name} & \colhead{(J2000)} & \colhead{(J2000)} & \colhead{$z$}  & \colhead{(km s$^{-1}$)}  & \colhead{arcmin / ($h_{70}^{-1}$ Mpc)} & \colhead{($R<2R_{\rm vir}$)} }
\startdata
A  & Cl1604+4304  & 241.097473 & 43.081150 & 0.898  & $703\pm110$ & 1.969 / 0.92 & 40 \nl
B  & Cl1604+4314  & 241.105050 & 43.239611 & 0.865  & $783\pm74$  & 2.261 / 1.05 & 62 \nl
C  & Cl1604+4316  & 241.031623 & 43.263130 & 0.935  & $304\pm36$  & 0.824 / 0.39 & 13 \nl
D  & Cl1604+4321  & 241.138651 & 43.353430 & 0.923  & $582\pm167$ & 1.594 / 0.75 & 60 \nl
F  & Cl1605+4322  & 241.213137 & 43.370908 & 0.936  & $543\pm220$ & 1.470 / 0.70 & 16 \nl
G  & Cl1604+4324  & 240.925080 & 43.401718 & 0.901  & $409\pm86$ & 1.143 / 0.53 & 15 \nl
\vspace*{-0.075in}
\enddata
\tabletypesize{\scriptsize}
\end{deluxetable*}
\end{center}

\vspace{-0.35in}
Early mid-IR studies of galaxy clusters carried out at 15 $\mu$m with
ISOCAM on the \emph{Infrared Space Telescope} succeeded in detecting a
substantial population of IR-luminous and presumably starbursting systems in
several clusters out to $z\sim0.5$ (see Metcalfe et al.~2005 for a review).  
More recently, observations at $24\mu$m with the Multiband Imaging Photometer
(MIPS) onboard the \emph{Spitzer Space Telescope} 
have allowed for detailed studies of star forming galaxies within 
clusters out to $z\sim1$.  Several studies have now detected a population of
Luminous Infrared Galaxies (LIRGs; $L_{\rm IR}>10^{11}$ $L_{\odot}$) with
unremarkable optical spectra in several distant clusters (Geach et al.~2006;
Marcillac et al.~2007; Dressler et al.~2009).  These galaxies often show
strong Balmer absorption and little or no [OII] emission due to severe dust
obscuration and were previously identified as possible buried starbursts by
Poggianti et al.~(1999) and Dressler et al.~(2004). 

Despite the identification of this population, the reasons for their
increased activity and the mechanisms that may trigger it remain unclear.
Since clusters in the early Universe have had less time to dynamically relax,
these galaxies may simply be recently accreted field galaxies whose
LIRG-level activity has yet to be quenched by the cluster environment.
Indeed LIRGs are far more prevalent in the field at $z\sim1$ than they are
locally (Perez-Gonzalez et al. 2005).  On the other hand, this activity may
be triggered by the physical stresses galaxies experience as they assemble
into denser environments.  Several mechanisms have been proposed which may
prompt starburst activity during cluster infall, including increased galaxy
mergers as a result of group compression (Barnes \& Hernquist 1991,
Struck 2006), tidal interactions (Moore et al.~1996), initial interactions
between galaxies and the intracluster medium (ICM; Evrard 1991), 
and the effects of a varying tidal field (Bekki 1999).

Thus far the study of several individual clusters have produced mixed results
regarding the nature of the IR-luminous cluster population.  On one hand,
multiple studies have found evidence for enhanced star formation in galaxies
as they approach the cluster environment for the first time.   
For example, Geach et al.~(2006) detected a sizable population of
LIRGs on the outskirts of two $z\sim0.5$ clusters and found evidence that the
level of obscured activity in each cluster is tied to the system's dynamical
state.  In addition, Marcillac et al.~(2007) determined that
the LIRGs detected in RXJ0152.7-1357 at $z=0.83$ are associated with infalling
galaxies, while Fadda et al.~(2008) reached the same conclusion after detecting
a starburst population in a filament feeding Abell 1787.  On the other hand,
Bai et al.~(2007) and Oemler et al.~(2009) find evidence of triggered
activity in galaxies that have recently passed through the center of MS1054
and Abell 851, respectively.  The role of galaxy interactions has also been
debated, with Oemler et al.~(2009) proposing that the merger of bound galaxy
pairs as they fall into the cluster core may lead to elevated star
formation, while Marcillac et al.~(2007) and Bai et
al.~(2007) find only mild evidence of merger driven activity in
RXJ0152.7-1357 and MS1054. 

If cluster-related processes play a role in triggering enhanced star
formation during cluster assembly, we can better constrain the mechanisms
responsible by studying starburst galaxies in clusters with a range of
dynamical states, levels of substructure and ICM densities.  To this end, we
have used $24\mu$m MIPS observations to identify and examine the starburst
population of six clusters and groups in the Cl1604 supercluster at
$z\sim0.9$.  The Cl1604 complex is the largest structure imaged by
\emph{Spitzer} at redshifts approaching unity and its constituent systems
are in varying stages of cluster formation and relaxation. 
These systems provide both a large sample of $24\mu$m-detected galaxies
and a diverse set of structures and local environments in which to study them.

In the following sections we make use of an extensive multiwavelength
dataset to examine the spatial distribution, IR properties, morphologies and
optical spectra of the $24\mu$m-detected galaxies in these systems in an
effort to constrain the mechanisms that may have prompted their increased
activity.  We also perform a detailed comparison between the physical
properties of these galaxies and those in lower density environments with
similar IR luminosities to determine if they are simply recently accreted
field galaxies.  We will show that while the $24\mu$m-bright cluster members
appear to be an infalling population, they are generally experiencing
burstier star formation activity than their counterparts in the field.
Furthermore, differences in their morphologies and optical colors preclude
them from simply being analogs of the IR-bright field population that have
yet to be quenched.   
  
We have organized this paper in the following manner:
in \S2 we describe the Cl1604 supercluster in greater detail, \S3 discusses
our multi-wavelength observations of the system, and \S4 outlines the details
of our sample selection and the method used to determine cluster membership
within the supercluster.  We then examine various properties of the
$24\mu$m-detected population.  We discuss their spatial distribution in \S5,
spectral properties and star formation activity in \S6, their morphologies in
\S7 and their optical colors in \S8.  We summarize our findings and their
implications in \S9 and our conclusions are presented in \S10.

\section{The Cl1604 Supercluster}

\begin{figure}[t]
\epsscale{1.1}
\plotone{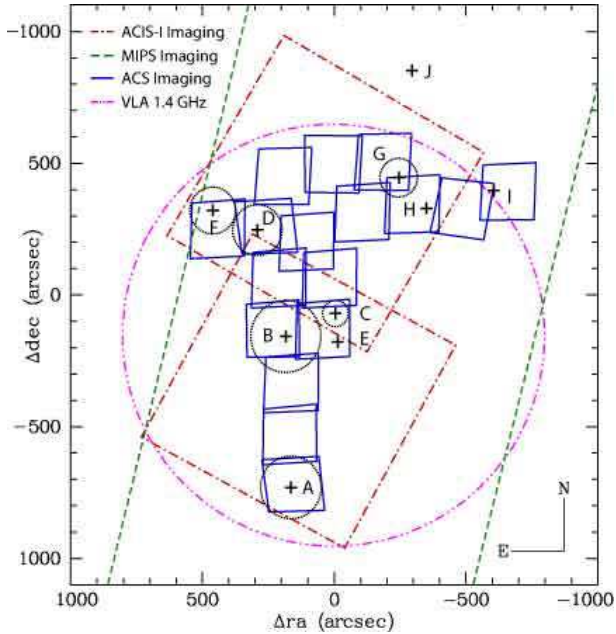}
\caption{Positions and FOVs of the 17 ACS pointings and single MIPS scan of
  the Cl1604 region.  Also shown is the FOV of the VLA and ACIS-I
  observations of the field.  The positions of the ten red-galaxy overdensities found by Gal et
  al. (2005, 2008) are marked and labeled following the naming convention of
  Gal et al.~(2005).  The virial radii of the six structures studied in
  this work are denoted by the black dotted circles. The figure is centered
  on $\alpha_{2000} = 16^{\rm h}04^{\rm m}7.6^{\rm s}$, $\delta_{2000} =
  +43^{\circ}17^{\prime}23^{\prime\prime}$. \label{fig-data_outline}}
\end{figure}

\begin{figure}[t]
\epsscale{1.2}
\plotone{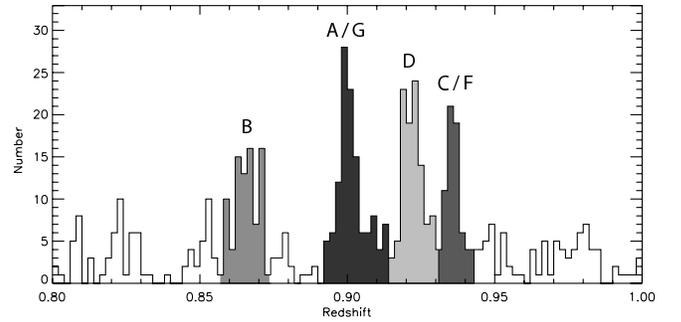}
\caption{Redshift distribution of the Cl1604 supercluster.\label{fig-zhist}}
\end{figure}

The Cl1604 supercluster is a high redshift, large-scale structure which
consists of eight spectroscopically confirmed galaxy clusters and groups, as
well as a rich network of filamentary structures.  The system spans roughly
10 $h_{70}^{-1}$ Mpc on the sky and 100 $h_{70}^{-1}$ Mpc in depth, with a
median redshift of $z\sim0.9$ (Gal \& Lubin 2004).  The velocity dispersions
of the system's constituent structures range from 300 km s$^{-1}$ to nearly
800 km s$^{-1}$ (Gal et al.~2005, 2008) and diffuse emission from the two
most massive clusters in the complex has been detected in our X-ray observations of
the field (Kocevski et al.~2009a).  The spatial distribution of the system's
eight clusters and groups are shown in Figure \ref{fig-data_outline} and the
system's redshift distribution is shown in Figure \ref{fig-zhist}. 

In this study we focus on the six best studied clusters and groups in the complex:
Cl1604+4304, Cl1604+4314, Cl1604+4321, Cl1604+4316, Cl1605+4322, and Cl1604+4324
(hereafter referred to as clusters A, B, and D and groups C, F, and G, respectively).
Various properties of these systems, including redshifts, velocity dispersions, 
virial radii, and the number of spectroscopically confirmed member galaxies,
are listed in Table \ref{tab-cl_prop}. 
  
Of these systems, cluster A is both the most massive and most relaxed structure in the complex.
It has a measured velocity dispersion of 703 km s$^{-1}$ (Gal et al.~2008)
and its ICM is X-ray detected with a bolometric luminosity of $L_{\rm X} =
 1.4\times10^{44}$ $h_{70}^{-2}$ erg s$^{-1}$ (extrapolated to $R_{200}$ assuming
a beta model with $\beta=2/3$; Kocevski et al. 2009a).
The cluster shows no signs of significant substructure and Gal et al.~(2008)
found no evidence of velocity segregation in its redshift distribution.
Cluster B is also X-ray detected with a luminosity slightly lower than that of
cluster A ($L_{\rm X} = 8.2\times10^{43}$ $h_{70}^{-2}$ erg s$^{-1}$), but
the system has a higher measured velocity dispersion (783 km s$^{-1}$).
Kocevski et al.~(2009a) found that the cluster departed from the $L_{\rm
  X}-\sigma$ relationship and concluded the system is not fully relaxed.
Cluster D has the third highest velocity dispersion in the supercluster (582 km s$^{-1}$), but
is not X-ray detected despite being relatively rich.  
Gal et al.~(2008) found significant levels of velocity segregation in the
cluster's redshift distribution, as well as an elongation of its member
galaxies that has been interpreted to be a galaxy filament feeding the cluster.  
The filament extends out over 2 $h_{70}^{-1}$ Mpc from the cluster center.
The system is the most dynamically active in the supercluster and is likely
in the process of accreting a substantial fraction of its future galaxy population.  Finally,
groups C, F, and G are poorer systems that are not directly detected in our
X-ray observations.  Their velocity dispersions range from 304 to 543 km
s$^{-1}$, consistent with rich groups, and each has at least a dozen
spectroscopically confirmed member galaxies. 

The supercluster has been extensively studied at a variety of wavelengths and
thus far over 1400 redshifts have been collected in the region,
resulting in spectra for 414 supercluster members (see \S\ref{opt-spect}).  The complex structure of
the supercluster, as mapped by our spectroscopic observations, is described in
Gal et al.~(2008), our X-ray observations of the system are
discussed in Kocevski et al.~(2009a), and our 20-cm
observations of the structure are described in Lubin et al.~(2010, in preparation).

\section{Observations and Data Reduction}

\subsection{IRAC and MIPS Imaging}
 
The entire Cl1604 supercluster was imaged at 3.6-8$\mu$m by IRAC and at
24$\mu$m by MIPS as part of the \emph{Spitzer} program GO-30455 (PI Lubin).
The area imaged by MIPS covers a region $\sim20^{\prime}\times60^{\prime}$ in
size and was observed in slow-scan mode with 8 scan legs and half-array
($148^{\prime\prime}$) cross-scan steps.  We repeated this pattern six times
to obtain a total exposure time of 1200 seconds per pixel for all but the
very edges of the resulting mosaic.   The IRAC observations cover a
$\sim20^{\prime}\times30^{\prime}$ subset of this area and consist of a grid
of $5\times6$  pointings with $260^{\prime\prime}$ separations.  At each
pointing we performed 36 medium-dithered, 30 second exposures, resulting in a
total exposure time of 1080 seconds per pixel throughout. 

The MIPS and IRAC data were processed using the standard \emph{Spitzer}
Science Center (SSC) reduction pipeline into individual Basic Calibrated Data
images.  The IRAC images were further processed using a modified version of
the SWIRE survey pipeline (Surace et al.~2005) to remove various instrumental
artifacts.  The resulting images were then coadded using the MOPEX software
and the final IRAC and MIPS mosaics have pixels scales of 0\farcs{6}
pixel$^{-1}$ and 1\farcs{2} pixel$^{-1}$, respectively

\begin{figure}[t]
\epsscale{1.15}
\plotone{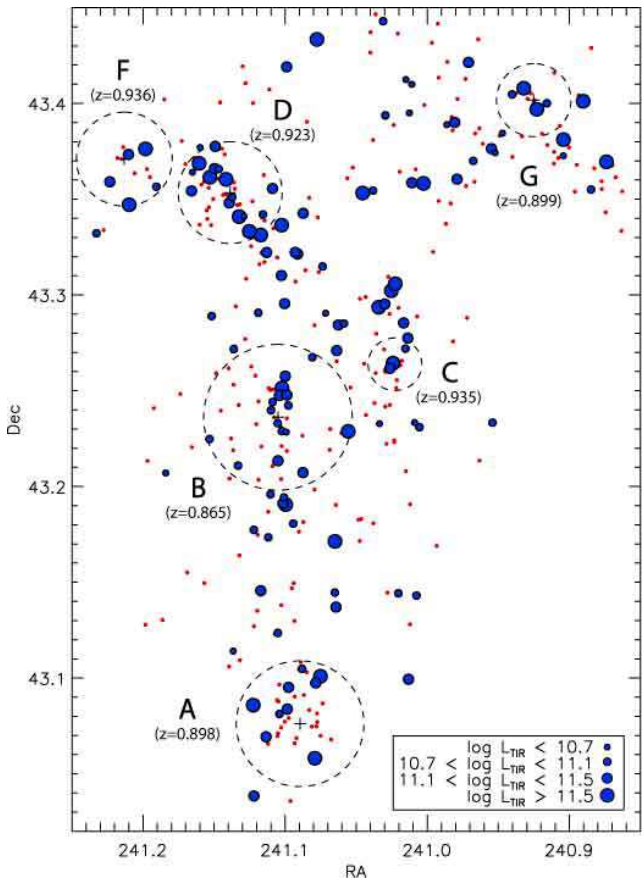}
\caption{Spatial distribution of galaxies within the Cl1604 supercluster
  ($0.84\le z\le0.96$).  The larger blue circles denote galaxies detected in
  the MIPS 24$\mu$m imaging, while the smaller red points denote those that
  are undetected.  The size of the blue circles scale with the total IR
  luminosity of the galaxies as indicated by the legend.  The dashed circles
  denote the virial radius of each system. \label{fig-spat_dist}} 
\end{figure}

Object detection on the IRAC imaging was carried out using SExtractor
primarily on the 3.6$\mu$m mosaic, but also on the 4.5$\mu m$ mosaic where
the two do not overlap.  Sources were selected if they had more than five contiguous
pixels above $1.8\sigma$ of the background fluctuations. 
Aperture photometry was carried out in all four IRAC
bands using dual-image mode in SExtractor with a fixed 1.9$^{\prime\prime}$
aperture.  We applied an aperture correction of 1.2 to account for the flux
outside our measurement aperture.  

Source detection in the 24$\mu$m mosaic was performed following the
methodology of the COSMOS survey described in Le Floch et al.~(2009).
Briefly, we use SExtractor to detect all sources at a significance of
$>1.8\sigma$ over an area of at least three pixels.  These positions are then
used as input to the DAOPHOT package (Stetson 1987) implemented within IRAF
(Tody 1986).  Bright, isolated sources are used to construct a single
empirical point spread function (PSF) for the entire mosaic.  
Using DAOPHOT, PSF fitting photometry is performed at each detected source
position and the flux enclosed within a 6$^{\prime\prime}$ aperture is
computed.  We then subtract the best-fit, scaled PSFs from the original
mosaic, and a second iteration of SExtractor and DAOPHOT is run to detect and
characterize any remaining sources.  The catalog from this second pass is
joined with the original catalog to produce the final source catalog.
Finally, we apply an aperture correction of 1.69.  With an exposure time of
1200 seconds and a moderate background level of 22.7 MJy/ster, we find our
3$\sigma$ point source sensitivity to be roughly 40$\mu$Jy, which is in good
agreement with estimates from the SSC's SENS-PET online tool.

\subsection{Optical Imaging}
The optical imaging of the supercluster used in this study consists of 17
pointings of the Advanced Camera for Surveys (ACS) on the \emph{Hubble Space
  Telescope (HST)}.  The ACS camera consists of two $2048\times4096$ CCDs
with a pixel scale of 0\farcs{05} pixel$^{-1}$, resulting in a
$\sim3^{\prime}\times3^{\prime}$ FOV.  The 17 pointing mosaic was designed to
image nine of the ten galaxy density peaks observed in the field of the
supercluster by Gal et al.~(2005, 2008).  Observations were taken in both the
F606W and F814W bands and consist of 15 pointings from GO-11003 (PI Lubin)
with effective exposure times of 1998 sec\footnote{With the exception of one
  pointing which lost guiding due to an incorrect attitude, resulting in a
  usable exposure time of 1505 sec and a gap in the mosaic.}
and 2 GTO pointings from G0-9919 (PI Ford) centered on clusters A
and D with effective exposure time of 4840 sec. Our average
integration time of 1998 sec resulted in completeness depths of $\sim26.5$
mag in each band.  

The ACS data were reduced using the pipeline developed by
the HST Archive Galaxy Gravitational Lens Survey (HAGGLES) team.  The
pipeline processing consists of calibrating the raw data using the best
reference files provided by the Hubble Space Telescope archive, subtracting
the background on each chip, iteratively determining the best shifts between
the dithered exposures using multiple calls to {\tt SExtractor} (Bertin \&
Arnouts 1996) and {\tt multidrizzle} (Koekemoer et al.~2002), aligning the
final drizzled image to the USNO-B1 catalog (Monet et al.~2002) and
resampling the images to a pixel scale of 0.03$^{\prime\prime}$ pixel$^{-1}$.
Object detection and photometry was carried out using {\tt SExtractor} in dual-image mode,
with the detection image being a weighted average of the F606W and F814W
images.  Detection parameters DETECT\_MINAREA and DETECT\_THRESH were set to
three pixels above $3\sigma$, while the deblending parameters DEBLEND\_NTHRESH
and DEBLEND\_MINCONT were set to 32 and 0.03, respectively.
An outline of the region covered by the ACS mosaic relative to the MIPS observations is
shown in Figure \ref{fig-data_outline}. Full details of the HAGGLES pipeline
can be found in Marshall et al.~(2010, in preparation). 

\subsection{Optical Spectroscopy}
\label{opt-spect}

\begin{figure*}[t]
\epsscale{1.15}
\plotone{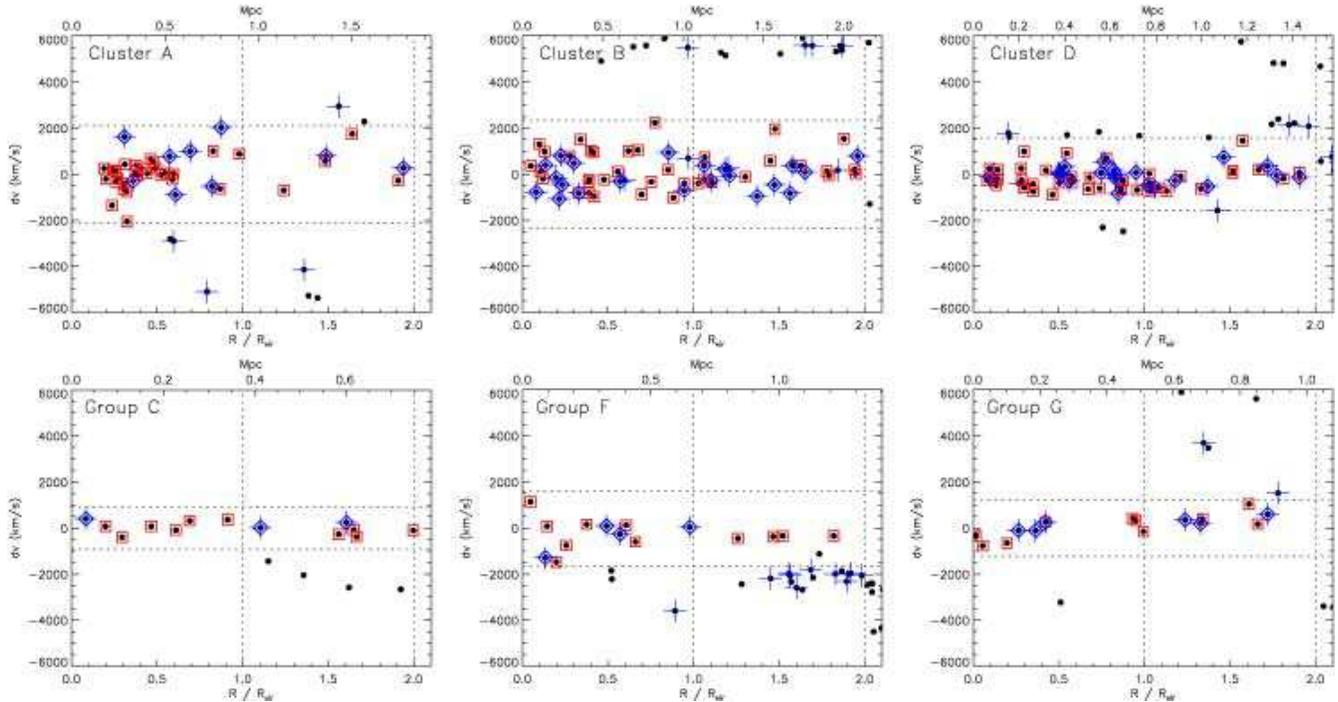}
\caption{The comoving velocity offset of galaxies from the median cluster/group
redshift versus cluster/group-centric distance.  A galaxy is considered a
member of a cluster or group if i.) its comoving velocity relative to the
median system redshift is less than three times the system's velocity
dispersion and ii.) it is located within two virial radii of the system
center.  All 24$\mu$m-detected galaxies are denoted by a blue cross, detected
cluster and group members are further highlighted with a blue diamond, and
undetected member galaxies are denoted by a red square. 
\label{fig-vel_dist}}
\end{figure*}

The Cl1604 supercluster has been extensively mapped using the
Low-Resolution Imaging Spectrograph (LRIS; Oke et al. 1995) and the
Deep Imaging Multi-object Spectrograph (DEIMOS; Faber et al.~2003) on
the \emph{Keck} 10-m telescopes (Oke, Postman, \& Lubin 1998; Postman, Lubin, \& Oke
1998; Lubin et al.~1998; Gal \& Lubin 2004; Gal et al.~2008). The
target selection, spectral reduction, and redshift measurements
are described in detail in Section 3 of Gal et al.~(2008). 
The resulting DEIMOS spectra have a resolution of $\sim$$1.7{\rm \AA}$ (68 km
s$^{-1}$) and typical spectral coverage from $6385{\rm \AA}$ to $9015{\rm
  \AA}$, while the LRIS resolution is $\sim$$7.8{\rm \AA}$ (330 km s$^{-1}$)
with spectral coverage from $5500{\rm \AA}$ to $9500{\rm \AA}$.

Our final spectroscopic catalog contains 1,638 unique objects. Redshifts
derived for these objects are given a spectroscopic quality, $Q_{spect}$,
between 1 and 4, where 1 indicates that a secure redshift could not be
determined, 2 is a redshift obtained from a single feature, 3 is a redshift
derived from at least one secure and one marginal feature, and 4 is assigned
to spectra with redshifts obtained from several high signal-to-noise
features. $Q_{spect} = -1$ is used for sources securely identified as stars.
In this sample, we find 140 stars and 1,138 extragalactic objects with
$Q_{spect} \ge 3$.  A total of 414 galaxies are in the nominal redshift range
of the supercluster, $0.84 \le z \le 0.96$.  The spatial distribution of
these galaxies is shown in Figure \ref{fig-spat_dist}.   

In addition to redshifts, we also use our optical spectra to measure the
equivalent widths (EW) of the [OII] and H$\delta$ features for spectral
classifications and to calculate SFRs from the [OII] line luminosity.
Due to the relatively low S/N of a single DEIMOS spectrum, bandpass
techniques are used for the EW measurements of individual galaxies.  
For consistency, we also adopt bandpass techniques to measure the absolute
luminosity of the [OII] emission line, making no corrections for the internal
extinction of each galaxy. Measurements are made adopting the standard
bandpasses of Fisher et al.~(1998) and calculated using the methodology
described in Lemaux et al.~(2010). For flux measurements, absolute
spectrophotometric calibration is achieved by using the methods described in
Lemaux et al.~(2010), adopting slit throughput of $\omega_{slit}=0.37$ for
all supercluster members. Star formation rates are derived from
[OII] line luminosities following the prescription of Kennicutt (1998).

\section{Identifying Supercluster Starbursts}

In order to find optical counterparts and assign redshifts to the 24$\mu$m
sources in the Cl1604 field, we performed a cross-correlation between our ACS
and MIPS selected samples using a nearest neighbor matching.  Employing a
match radius of $1\farcs{44}$, we found 283 MIPS sources matched to galaxies
with $Q_{spect} \ge 3$ redshifts.  Of these, 126 have redshifts between
$0.84\le z \le0.96$ and fall within the boundaries of the supercluster.  The
location of these galaxies within the supercluster is shown in Figure
\ref{fig-spat_dist}.  Using the average surface density of 24$\mu$m sources
with $f_{24} > 40$ $\mu$Jy (measured throughout the supercluster) and a match
radius of $1\farcs{44}$, we calculate a spurious match probability of 1.3\%
from the P-statistic (Lilly et al.~1999).  We therefore expect 2 of 126
matches between a supercluster member and a 24$\mu$m source to be spurious.

\subsection{Determining Cluster Membership}

To determine the host cluster or group of each 24$\mu$m-detected galaxy within the
complex structure of the supercluster, we use the galaxy's position and
velocity offset relative to the systemic position and redshift of each cluster/group.  
Creating cluster and group specific subsamples in this manner will allow us to examine the
large-scale environments of the starburst population in terms of the
properties of their host systems.  We consider a galaxy associated to a
specific cluster or group if 
i.) its comoving velocity relative to the median cluster/group redshift is
less than three times the system's velocity dispersion, $\sigma_{\rm v}$, and
ii.) it is located within two projected virial radii, $R_{\rm vir}$, of the
cluster/group center.  This latter constraint is motivated not only by a
desire to sample a full range of environments around each system, but also
due to a large population of infalling galaxies observed near multiple Cl1604
clusters that extend beyond $R_{\rm vir}$; this is discussed further below.

We determined $R_{\rm vir}$ for each cluster and group using the system's measured
$\sigma_{\rm v}$ and the relation $R_{\rm vir} = R_{200}/1.14$ (Biviano et
al.~2006; Poggianti et al.~2009), where $R_{200}$ is the radius at which the
density of the cluster is 200 times that of the critical density of the
Universe.  We measured $\sigma_{\rm v}$ for each system using an iterative
clipping procedure that is described in detail in Gal et al.~(2008), while
$R_{200}$ was in turn calculated using the equation 

\begin{equation}
  R_{200} = \frac{\sqrt{3}\sigma_{\rm v}}{10 H(z)}.
\end{equation}

\noindent The derived values of $\sigma_{\rm v}$ and $R_{\rm vir}$ for the
Cl1604 clusters and groups are listed in Table \ref{tab-cl_prop}.  

The result of applying the two conditions $|dv|<3\sigma$ and $R<2R_{\rm vir}$
for both 24$\mu$m-detected and undetected galaxies near clusters A, B, and D
and groups C, F, and G are shown on the top and bottom panels of Figure
\ref{fig-vel_dist}, respectively.
The differing dynamical states of these clusters is readily apparent. 
In the case of the relaxed cluster A, a large majority of the system's spectroscopically confirmed members fall
within $R_{\rm vir}$, while in the dynamically unrelaxed clusters B and D, a substantial
population are found at greater radii.  
We interpret this as evidence that both systems are actively accreting a
considerable fraction of their galaxy populations.

Using the above selection criteria, we have associated a total of 71
of the 126 24$\mu$m detected supercluster members with a
specific cluster or group in the Cl1604 complex.  The remaining galaxies that fall within the
redshift boundaries of the supercluster are found in loose groups, filamentary
or sheet-like structures near these systems.  For the remainder of the paper,
we will refer to galaxies associated with clusters A, B, and D as
\emph{cluster members} and those associated with groups C, F, and G as
\emph{group members}, while the remaining galaxies will be referred to as the
\emph{supercluster field} sample (or simply the \emph{superfield} population for short).

\subsection{Total IR Luminosity \& SFRs}

\begin{figure}[t]
\epsscale{1.15}
\plotone{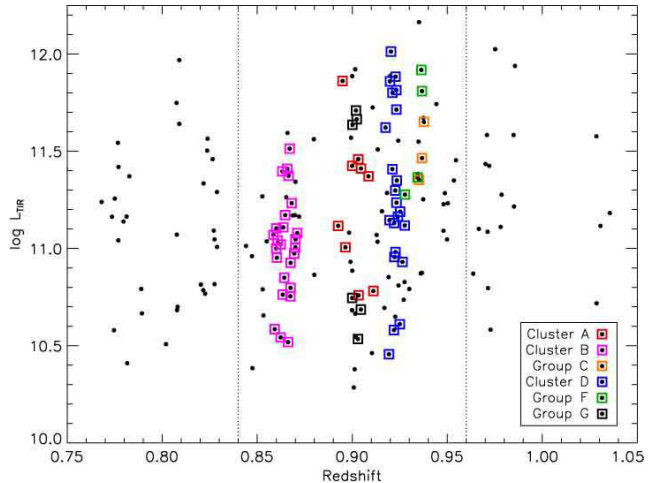}
\caption{Total IR luminosity versus redshift for galaxies in the field of the
  Cl1604 supercluster.  The vertical dotted lines denote the traditional
  redshift boundaries of the supercluster.  Specific cluster and group
  members are labelled according to the legend. \label{fig-z_ltir}}
\end{figure}

To determine the level of obscured activity occurring in each 24$\mu$m source
with a measured redshift, we calculate the galaxy's total IR luminosity
($L_{\rm TIR}$; 8-1000 $\mu$m) from its monochromatic 24$\mu$m flux density
using the recipe of Chary \& Elbaz (2001) and the synthetic spectra of Chary
\& Elbaz (2001) and Dale \& Helou (2002).  
These spectral templates model the 8-1000 $\mu$m spectral energy distribution
(SEDs) of galaxies with a wide range of IR luminosities and by determining
which template best reproduces the 24$\mu$m flux of a given galaxy at a given
redshift, we can establish the galaxy's $L_{\rm TIR}$.  
This was done for each galaxy by redshifting the templates to the galaxy
redshift, converting them from luminosity to flux per unit wavelength, and
convolving the SED with the MIPS 24$\mu$m response curve.  The template that
best reproduced the 24$\mu$m flux of the galaxy was then normalized to this
flux and integrated to obtain $L_{\rm TIR}$.  

We used the derived $L_{\rm TIR}$ values to obtain the level of star
formation occurring in each galaxy by way of the $L_{\rm TIR}$-SFR relation from Kennicutt (1998):

\begin{equation}
{\rm SFR} \hspace{0.05in} (M_{\odot} \hspace{0.05in} {\rm yr}^{-1}) = 4.5\times10^{44} L_{\rm
  TIR} \hspace{0.05in} ({\rm ergs \hspace{0.05in} s}^{-1})
\end{equation}

\vspace{0.2in}
\noindent The distribution of $L_{\rm TIR}$ as a function of redshift for supercluster
members is shown in Figure \ref{fig-z_ltir}.  Our $3\sigma$ flux limit
of 40 $\mu$Jy corresponds to a luminosity of $3\times10^{10}$ $L_{\odot}$ at $z=0.9$, the median redshift
of the supercluster.  This luminosity in turn corresponds to a SFR of 5.2 M$_{\odot}$
yr$^{-1}$.  Therefore our MIPS observations allow us to sample the
luminous end of the IR normal ($L_{\rm
  TIR} < 10^{11}$ $L_{\odot}$), LIRG ($L_{\rm TIR} = 10^{11-12}$
$L_{\odot}$), and Ultra LIRG (ULIRG; $L_{\rm TIR} > 10^{12}$ $L_{\odot}$)
populations in the Cl1604 systems.  Only two supercluster members have luminosities
consistent with ULIRGs, while the majority (64.3\%) have LIRG-like
luminosities.  The remaining 34.1\% fall in the IR normal regime.   
For simplicity we will often refer to galaxies in all three luminosity
classes collectively as the starburst population.  In \S6 we justify this by
demonstrating that a significant fraction of these galaxies are experiencing
bursty episodes of star formation activity, especially those within the
cluster and group environments. 

\begin{figure}[t]
\epsscale{1.15}
\plotone{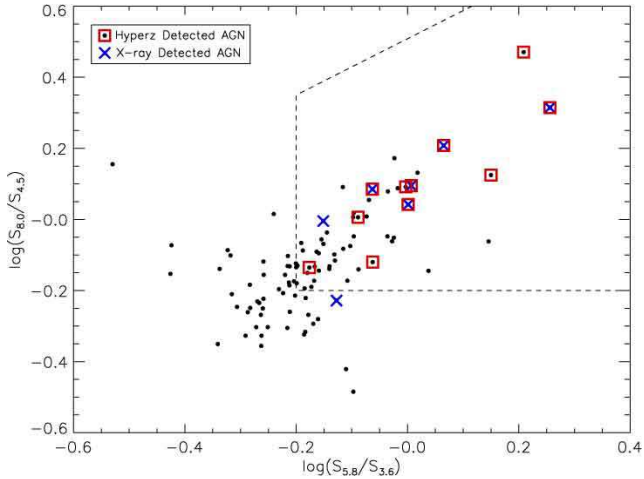}
\caption{IRAC color-color plot based on that of Lacy et al.~(2004).  All
  cluster and group members are shown in black, AGN detected in our
  \emph{Chandra} X-ray imaging are denoted by blue crosses, and AGN
  selected by our Hyperz SED fitting technique are denoted by red squares.
  \label{fig-hyperz}}
\end{figure}

\subsection{AGN Contamination}

\begin{figure*}[t]
\epsscale{1.1}
\plotone{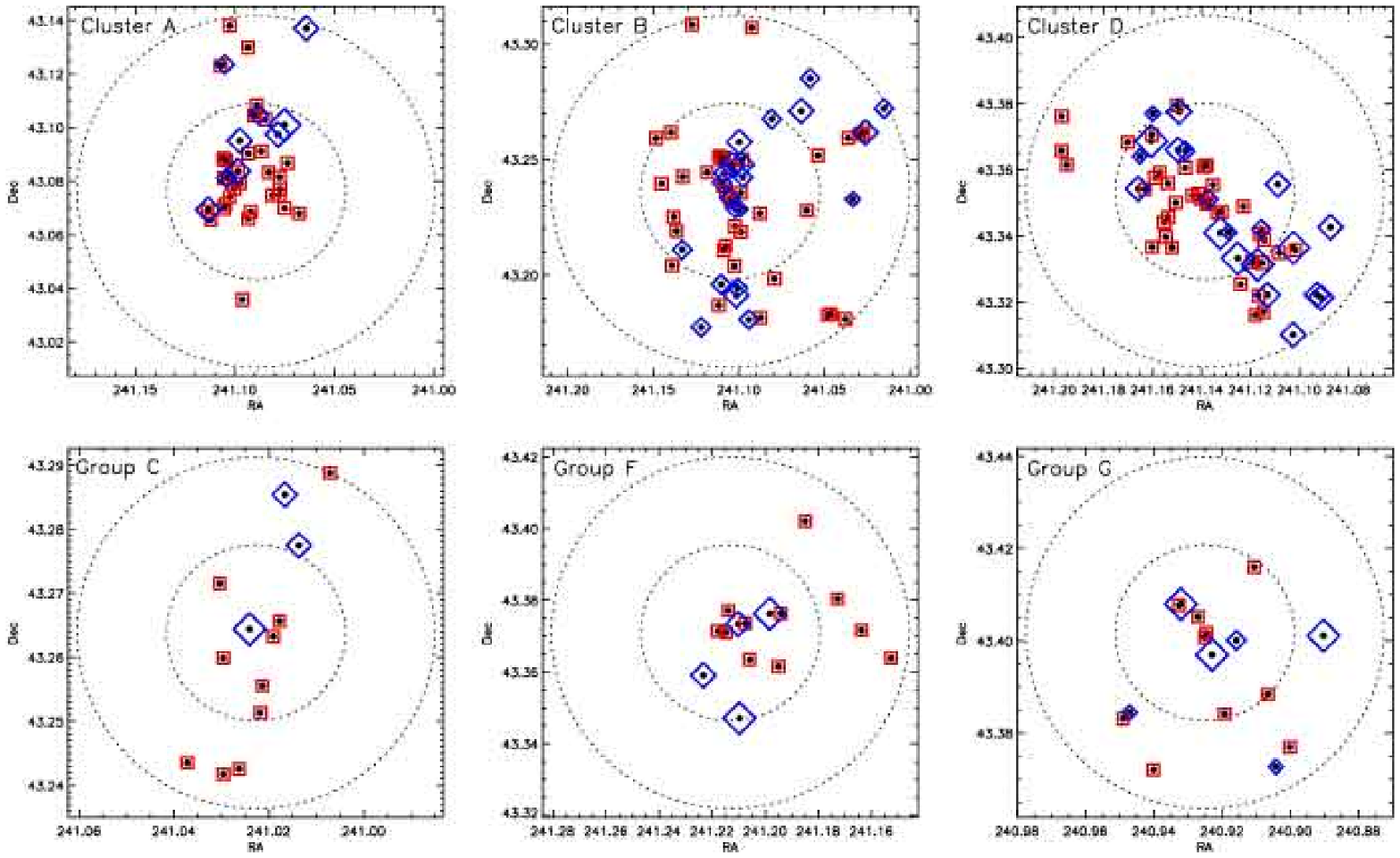}
\caption{Spatial distribution of 24$\mu$m-detected (\emph{blue diamonds}) and
  undetected (\emph{red squares}) cluster and group members.  The inner circles
  denote the virial radius of each system, while the outer circles are twice
  this distance.  The symbol size of the 24$\mu$m-detected galaxies scale
  with total IR luminosity in the same manner as shown in Figure 3. 
  We find a greater fraction of 24$\mu$m-bright galaxies in the more
  dynamically unrelaxed clusters B and D. 
   \label{fig-radec_mem24}}
\end{figure*}

The mid-IR flux of galaxies can originate from both obscured star formation
and the reprocessing of AGN related emission by dust. In this section we discuss
the identification of galaxies whose mid-IR flux is dominated by the latter
in order to exclude them from any analysis of the supercluster's starburst
population.  To do this, we have fit starburst and AGN spectral templates to
our ground-based optical, IRAC and MIPS photometry using the Hyperz
photometric redshift software (Bolzonella et al.~2000).  The templates we use
for these fits are a subset of the SWIRE template library (e.g. Polletta et
al.~2007) and include five spiral (S0, Sa-Sd), three starburst and four AGN
templates. The starburst SEDs are those of NGC 6090, M82 and Arp200, while
the AGN templates are a combination of models, broadband photometry and
composite spectra of local AGN and correspond to Seyfert 1.8, Seyfert 2, QSO1
and QSO2 SEDs.  We adopt a Calzetti reddening prescription (Calzetti et
al.~2000) with up to a magnitude of optical extinction to account for heavily
obscured sources. 

We carried out the Hyperz SEDs fitting on 98 supercluster members
detected at 24$\mu$m and all four IRAC bands (out of a total 126 detected at
24$\mu$m).  Of this sample, three were best fit by QSO SEDs and another eight
 were well matched to Seyfert SEDs.  The location of these galaxies in IRAC
color-color space is shown in Figure \ref{fig-hyperz}.  All of the Hyperz
selected AGN fall within the color selection wedge of Lacy et al.~(2004) and
they generally follow the color trend expected for sources with a negative
power-law spectral slope (i.e.,~Donely et al.~2008).   Also shown in Figure
\ref{fig-hyperz} are the colors of X-ray luminous AGN found in the
supercluster through our \emph{Chandra} observations of the system (Kocevski
et al.~2009a,b).  There are a total of nine X-ray detected AGN in the
supercluster with Seyfert-like luminosities.  Seven of these are detected at
24$\mu$m and in all four IRAC bands, of which five were selected as AGN by our
Hyperz fitting.  The outstanding two were found to have a significant
starburst component and are most likely starburst/AGN blends.  In order to be
conservative in our AGN selection, we have not employed hybrid starburst-AGN
templates which would have selected these sources as AGN, but plan to in a
future study of the obscured AGN population within the supercluster.

The final sample of AGN excluded from our analysis consists of 13
sources (five IR/X-ray selected, six IR selected, and two X-ray selected)
out of 98 potential hosts, resulting in a roughly 13\% contamination.  An
additional 28 sources were not screened using Hyperz because they lacked a
detection in one or more of the IRAC bands.  Given our contamination rate, we
expect 4 of these galaxies to host an AGN.  Since these make up less than 4\%
of our final sample, we do not believe their presence will significantly
alter our results. 

Of the 13 AGN found using Hyperz, 7 galaxies are part of our cluster/group
sample, while the remaining 6 are part of the supercluster field. The only
two cluster/group members with ULIRG-level luminosities are among the three
galaxies best-fit by the QSO SEDs and are located near the center of Cluster D.
With these AGN excluded, our final sample of 24$\mu$m sources consists of 50
cluster, 13 group and 50 field galaxies.   In the following sections we take
a detailed look at the various properties of these galaxies.

\section{Spatial Distribution}

The spatial distribution of 24$\mu$m-detected galaxies over the entire Cl1604 
supercluster is shown Figure \ref{fig-spat_dist}, while those associated with
specific clusters and groups in the complex are shown in Figure \ref{fig-radec_mem24}. 
The distribution of these galaxies provides several immediate insights.

\begin{center}
\begin{deluxetable}{lccc}
\tabletypesize{\scriptsize}
\tablewidth{0pt}
\tablecaption{Projected Density of 24$\mu$m-Detected Galaxies\label{tab-densities}}
\tablecolumns{4}
\tablehead{\colhead{}  & \colhead{} & \colhead{$N_{\rm 24\mu m}$ } & \colhead{$n_{\rm 24\mu m}$}  \\
           \colhead{Region} & \colhead{$N_{\rm 24\mu m}$} & \colhead{Corrected} & \colhead{(gal/arcmin$^{2}$)}}
\startdata
Cluster A      &  9 & 15.0 & 1.235$^{+0.407}_{-0.314}$ \nl
Cluster B      & 15 & 25.7 & 1.568$^{+0.376}_{-0.309}$ \nl
Cluster D      & 14 & 18.7 & 2.338$^{+0.680}_{-0.541}$ \nl
Cl1604 Field 1 &  9 & 15.2 & $0.691^{+0.113}_{-0.087}$ \nl
Cl1604 Field 2 & 13 & 26.5 & $1.046^{+.130}_{-0.107}$ \nl
\hline
Avg Cluster    & 38 & 59.4 & 1.625$^{+0.239}_{-0.210}$ \nl
Avg Field   & 33 & 63.5 & 0.881$^{+0.159}_{-0.137}$  \nl
\vspace*{-0.075in}
\enddata
\tabletypesize{\scriptsize}
\tablecomments{Density of sources with F814W$<23.5$ and in the
  redshift range $0.84\le z \le0.96$.}
\end{deluxetable}
\end{center}

\vspace{-0.35in}
\subsection{Increased Activity in Unrelaxed Clusters}

First, we find increased activity in the dynamically unrelaxed clusters B and
D relative to the more massive and relaxed cluster A.  The fraction of
cluster members detected at 24$\mu$m in the latter is 22.5\%, while in
cluster B and D this increases to 36.2\% and 34.5\%, respectively.  These
increased fractions are inconsistent with the cluster A fraction at roughly
the 90\% level.  This finding suggests that the dynamical state of these
clusters is directly related to the level of dust-enshrouded star
formation occurring in their member galaxies.  

In the case of cluster D, the 24$\mu$m-detected galaxies are preferentially
located either within the filament connected to the system or aligned with
its infall direction.  This is one of the clearest indications from our data
that enhanced star formation within these systems is associated with actively
infalling or recently accreted galaxies.  This finding is similar to the
results of Fadda et al.~(2008), who detected an overdensity of starburst
galaxies in two filaments feeding Abell 1763.  The authors suggest the
filament environment is a favorable setting for the onset of starburst
activity; a conclusion that our observations support.

We have performed an extensive test to confirm that the increased activity we
find among the unrelaxed clusters is not due to variations in our
spectroscopic sampling.  After carefully correcting for our spectroscopic
incompleteness, we find the density of 24$\mu$m sources in the redshift range
$0.84\le z\le 0.96$ is significantly greater near clusters B and D relative
to both cluster A and the general field population.  We performed this
calculation in the following manner.  To correct for our spectroscopic
incompleteness, we binned our spectroscopic sample of 1278 objects in
observed color (F606W-F814W) and magnitude (F814W) space and calculated the
ratio of galaxies confirmed to lie within $0.84\le z\le 0.96$ to those
outside this range for each color-magnitude bin.  We then used these ratios
to determine the probability that a given 24$\mu$m source without a redshift
would fall within the supercluster redshift range had it been targeted for
spectroscopic observations.   These probabilities are then used as weights,
which are summed to estimate the number of 24$\mu$m-detected members missed
in any given region within the supercluster.  In carrying out this analysis
we chose our color and magnitude bins to be 0.5 and 1.0 magnitudes wide,
respectively, in order to ensure adequate number statistics in each bin and
we have adopted a magnitude limit of F814W$<23.5$.   

With an estimate of our spectroscopic incompleteness, we chose three cluster
environments ($R<R_{\rm vir}$ in systems A, B, and D) and two field
environments in which to compare the number density of 24$\mu$m sources.  
The field regions have been chosen to avoid the cluster and group cores, but
to remain within our ACS imaging and the regions well sampled by our
spectroscopy.   Although these regions will undoubtedly be denser than the
nominal field well outside the Cl1604 supercluster, we have decided to compare against
them as opposed to field surveys since the 24$\mu$m flux limit and
our spectroscopic and ACS coverage in these regions are identical to that of the
cluster cores.  This allows for a fairly straightforward comparison of the source
densities in these different environments.

The corrected number density of 24$\mu$m sources in each region,
integrated over the redshift range of $0.84\le z\le 0.96$ are listed in Table
\ref{tab-densities}.  We find an excess of 24$\mu$m sources within the
virial radius of each cluster relative to the supercluster field.  In the
three regions encompassing the cluster centers, the average projected density
of dusty, star forming galaxies is nearly twice that of the adjacent field.
This overdensity is detected with a statistical significance of just over
$3\sigma$.  Furthermore, the excess of 24$\mu$m-detected galaxies in each
cluster appears well correlated with the system's dynamical state.  We find a
nearly two-fold increase in the density of sources in the dynamically
unrelaxed cluster D relative to the more relaxed cluster A.   
This is consistent with our earlier finding that the two dynamically
disturbed clusters have a greater fraction of 24$\mu$m-detected members.

It is important to note that the overdensity of 24$\mu$m sources found in
clusters B and D cannot simply be a result of the higher overall density of
galaxies in these systems as a similar excess is not observed in cluster A.
In fact, cluster A is more massive and optically richer than either of its
counterparts. We would have, therefore, expected to find a greater excess in
that system if the density of 24$\mu$m sources simply scaled with the general
galaxy population and this is not the case.
 
Our findings are similar to those of Geach et al.~(2006),
who detected a substantial population of 24$\mu$m-detected galaxies in Cl0024+16 and
relatively few in the more massive and relaxed cluster MS0451-03.
Given that dynamically unrelaxed clusters are typically undergoing active
assembly, the increased activity observed in these systems further suggests that the
starburst population consists of recently accreted field galaxies whose
activity is either triggered directly prior to cluster infall or by cluster-specific
processes during infall.

\begin{figure}[t]
\epsscale{1.15}
\plotone{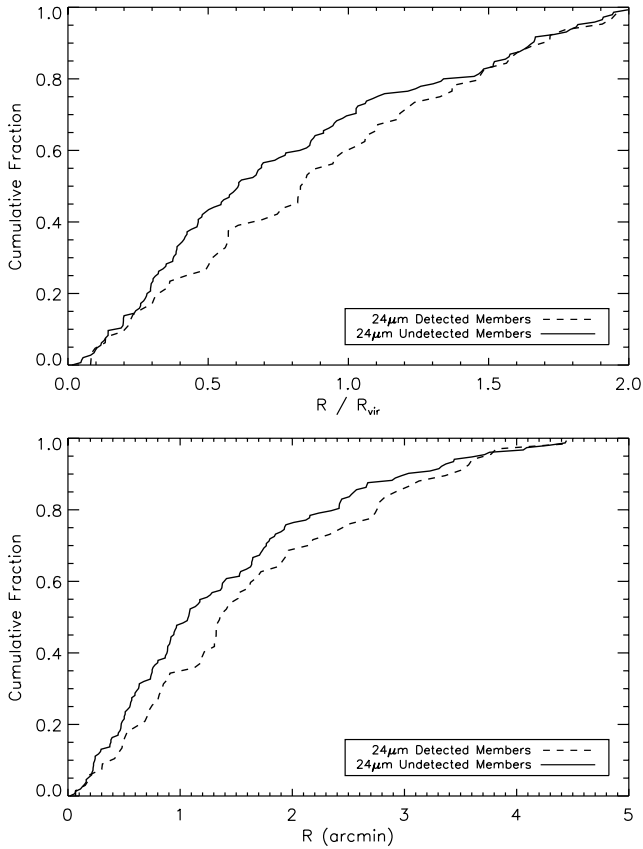}
\caption{Cumulative distribution of 24$\mu$m-detected and undetected galaxies
  versus distance from the cluster/group centers; shown both in terms of
  arcminutes (\emph{bottom panel}) and normalized to each system's virial radius
  (\emph{top panel}).  We find that the 24$\mu$m-detected galaxies are less
  centrally concentrated than their undetected counterparts when summed over
  all of the Cl1604 clusters and groups  \label{fig-clcent_dist}} 
\end{figure}

\subsection{Radial Distribution}

\begin{figure}[t]
\epsscale{1.15}
\plotone{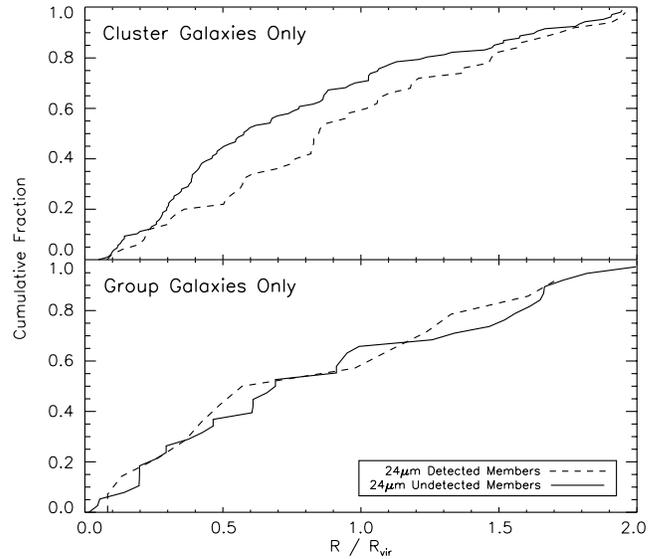}
\caption{Cumulative distribution of 24$\mu$m-detected and undetected galaxies
  versus distance for the cluster (\emph{top panel}) and group (\emph{bottom
    panel}) samples separately.  Distances are normalized to each system's
  virial radius.  While the detected cluster members are less centrally
  concentrated than their undetected counterparts, this is not the case with
  the group population.   \label{fig-clcent_dist2}}
\end{figure}

The second insight provided by the spatial distribution of the 24$\mu$m
population is that they are less centrally concentrated than their
undetected counterparts when summed over all of the Cl1604 clusters and groups
(systems A through G).  This can be seen in Figure \ref{fig-clcent_dist},
which shows the cumulative distribution of 24$\mu$m-detected and undetected
member galaxies as a function of cluster/group-centric distance.  The distribution
is shown both in terms of arcminutes to the center of each galaxy's host system
and distances normalized to each system's $R_{\rm vir}$.  Employing a KS test
on the normalized distribution, we calculate only a 14.3\% probability that
the 24$\mu$m detected sample is drawn from the same parent distribution as
the undetected population\footnote{This decreases to 7.9\% if we do not
  normalize the cluster-centric distances.}.   
When we split the 24$\mu$m sample into cluster (A, B, and D) and group (C, F,
and G) subsamples, we find that this result is largely driven by the cluster
population.  This can be seen in Figure \ref{fig-clcent_dist2}, which shows
the same cumulative distribution, but for the cluster and group subsamples
separately.  While the  24$\mu$m cluster members are less centrally
concentrated than the undetected cluster members, this is not the case with the
group population.  A KS test on each distribution separately confirms this
dichotomy.  We find an 87.4\% chance that the 24$\mu$m group members are
drawn from the same distribution as the undetected group members and only a
3.5\% probability of this being true for the 24$\mu$m cluster population. 
Undoubtedly the radial distribution of the cluster members is heavily
influenced by the large number of 24$\mu$m-bright galaxies in the filament
extending outward from cluster D.

This difference in the spatial concentration of group and cluster
24$\mu$m galaxies was also recently observed by Krick et al.~(2009),
who measured a less concentrated distribution of star forming galaxies in a
massive, evolved cluster at $z=1$ than in two dynamically younger systems at
the same redshift.  These results imply that while starburst activity is
found throughout the group population, it is predominantly found among
infalling galaxies in the cluster environments.

\section{Spectral Properties and Star Formation Activity}

\subsection{Spectral Classification}

At the redshift of the Cl1604 systems the spectral window of our DEIMOS
observations encompass the [OII] doublet at $3727$\AA$ $ and the H$\delta$
absorption line at $4101$\AA$ $, which we can use to probe the dominant mode
of star formation activity (i.e.,~bursty versus continuous) occurring in the
24$\mu$m-detected galaxies.  Much has been written in the literature
regarding the ability of these features to provide insight into the recent
star formation histories of galaxies (see Oemler et al.~2009 for a recent
review).  We briefly discuss the utility of these lines below, before
presenting our spectral classification of the 24$\mu$m-bright population. 

Due to the lifetime of the stars that give rise to each feature, namely O and
B-type stars for [OII] and longer-lived A-type stars for H$\delta$, these lines
provide a measure of the average SFR of a galaxy over different timescales,
with [OII] measuring current activity ($\tau\sim10^{7}$ yr) and H$\delta$ reflecting
more extended activity ($\tau\sim10^{9}$ yr).  As a result, the strengths of
these lines, in the absence of significant dust extinction, are well understood 
for different types of activity.  In galaxies undergoing normal, continuous
star formation, the EW of both [OII] and H$\delta$ increases with increasing
activity until the SFR is high enough that H$\delta$ emission from new stars
begins to infill the H$\delta$ absorption line, leading to a overall decrease in the
line's EW.  On the other hand, in bursty conditions, when an epoch of high star formation
activity is followed by one with significantly less, the EW of H$\delta$
increases dramatically as there is insufficient H$\delta$ emission from the
current generation of stars to infill the H$\delta$ absorption produced by the
previous generation.  Galaxies that have experienced a rapid reduction in
their activity, i.e., post-quenched systems, show strong H$\delta$ absorption
for the same reasons, except that they also have little or no [OII] emission.

\vspace*{-0.2in}
\begin{center}
\begin{deluxetable}{cccc}
\tabletypesize{\scriptsize}
\tablewidth{0pt}
\tablecaption{Spectral Properties of 24$\mu$m-Detected Galaxies\label{tab-spect}}
\tablecolumns{4}
\tablehead{\colhead{Spectral}  & \colhead{Cluster/Group} & \colhead{Superfield} & \colhead{}   \\
           \colhead{Class} & \colhead{($R<2R_{\rm vir}$)} &
           \colhead{($R>2R_{\rm vir}$)} & \colhead{Comment} }
\startdata
e(a)   & 28 (47.5\%)  & 13 (32.5\%)  & Obscured Starburst \nl
e(b)   & 10 (17.0\%)  & 04 (10.3\%)  & Starburst \nl
e(c)   & 19 (32.2\%)  & 23 (57.5\%)  & Continuous SFR \nl
k+a    &  02 (03.4\%)  & 00 (00.0\%)  & Post-Starburst \nl
\vspace*{-0.075in}
\enddata
\tabletypesize{\scriptsize}
\end{deluxetable}
\end{center}

\vspace{-0.1in}
Using these characteristics, Dressler et al.~(1999) devised a set of
spectral classes based on the EW of [OII] emission and H$\delta$ absorption
that reflect different modes of star formation activity.  In this classification
scheme e(a) and e(b) galaxies are ongoing starbursts with varying levels of
dust obscuration.  While both classes exhibit strong Balmer absorption
[EW(H$\delta$) $>$ 4\AA], only moderate [OII] emission (EW[OII] $>$ -40 \AA)
is visible in e(a)'s, while e(b)'s show strong [OII] emission (EW[OII] $<$
-40 \AA).  On the other hand, the spectra of e(c) galaxies are consistent
with normal star formation, having moderate levels of both [OII] emission
(EW[OII] $>$-40 \AA) and Balmer absorption [EW(H$\delta$) $<$ 4\AA]. 
Finally k+a, or post-starburst systems, are characterized as having
strong Balmer absorption features superimposed on an older K-star spectrum and
lacking [OII] emission.
 
Using this spectral classification scheme we have classified the
24$\mu$m-detected cluster, group and field galaxies based on the EW of [OII]
emission and H$\delta$ absorption measured in their spectra.  
Out of the 113 non-AGN, 24$\mu$m-bright supercluster members, 101 have spectra
with adequate S/N for this analysis.  The EWs were measured as described in
\S3.4 and a line was considered detected if its measured EW is greater than
three times the error associated with that measurement.
The results of our analysis are listed in Table \ref{tab-spect}.  In general
we find that the 24$\mu$m-detected cluster and group members predominantly
have e(a) and e(c) type spectra, which indicates they are comprised of a mix
of obscured starbursts and galaxies undergoing normal, continuous star
formation.  We find similar results in the superfield sample, with the
notable difference being a reversal in the e(a) to e(c) ratio.  
A greater fraction of the cluster and group galaxies appear to be undergoing
dusty starburst activity compared to the field sample (47.5\% versus 32.5\%),
where a majority have spectra consistent with normal star formation (57.5\%).
The reversal in the e(a) to e(c) ratio indicates that star forming galaxies
in these denser environments are, on average, experiencing burstier activity
than their counterparts in the field. 

\begin{figure}
\epsscale{1.15}
\plotone{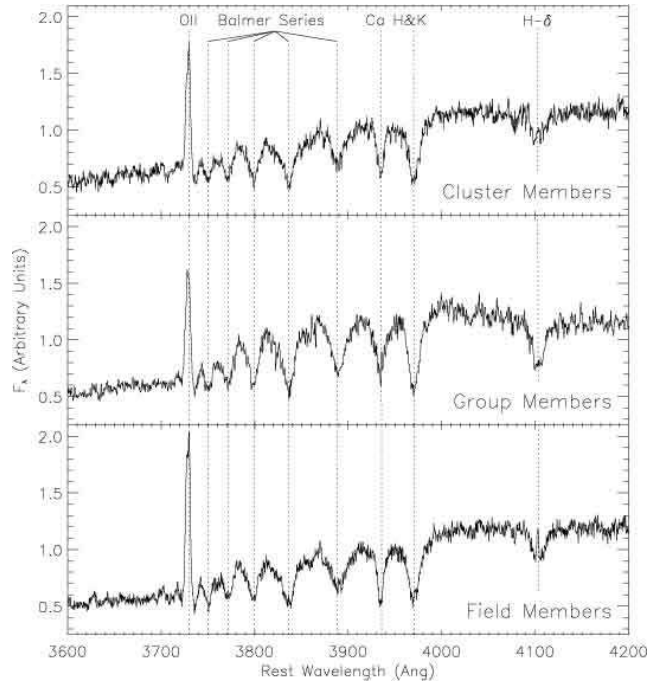}
\caption{Stacked rest-frame composite spectra of the 24$\mu$m-detected
  cluster, group and superfield subsamples.  \label{fig-spectra}}
\vspace{-0.1in}
\end{figure}

\subsection{An Excess of Starburst Activity}

Since this result is based on measurements taken from individual spectra that
typically have only moderate-S/N, we have stacked the spectra of the
24$\mu$m-detected galaxies as a function of environment to produce high-S/N
composite spectra in order to confirm our findings.
These co-added spectra can be used to gauge the average spectral properties of the
stacked subsamples to greater accuracy than is possible using single galaxy
spectra (e.g. Dressler et al.~2004).  We will use these properties to compare
the dominant mode of star formation (i.e.,~bursty versus continuous)
occurring in the cluster, group and superfield populations.

To produce the composites, the spectra of galaxies in each region are
normalized and averaged using an inverse variance weighting.  We find that
the resulting spectra do not vary significantly depending on whether we use
a unit- or (optical) luminosity- weighted average, especially for the most
luminous 24$\mu$m sources.  

In total the group, cluster, superfield subsamples included 8, 30, and 47
24$\mu$m-detected galaxies, respectively.  These totals decrease to 7, 20,
and 28, respectively, when we only consider galaxies with LIRG-level
luminosities ($L_{\rm TIR}>10^{11}$ $L_{\odot}$).  For each stack, the
cluster and group samples were restricted to galaxies within the virial
radius of each system.  The resulting composite spectra for the 24$\mu$m
population of each region are shown in Figure \ref{fig-spectra}.   

\begin{figure}
\epsscale{1.1}
\plotone{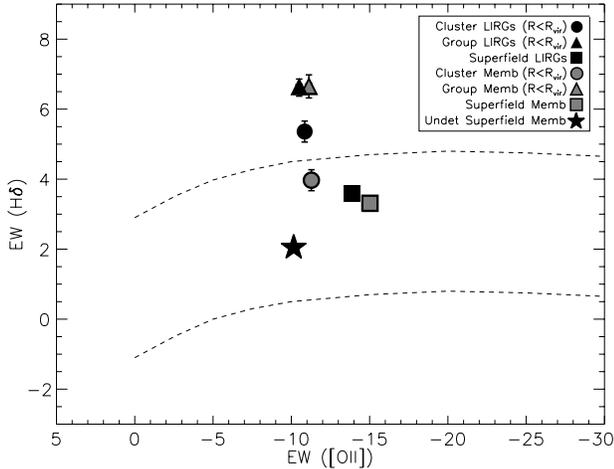}
\caption{The average EW of $H\delta$ versus that of [OII] measured in the
  cluster, group and superfield composite spectra.  Results are shown for both
  the entire 24$\mu$m-detected sample (\emph{grey symbols}) and only those with
  LIRG-level luminosities (\emph{black symbols}).  For comparison the
  undetected superfield sample is also shown (\emph{black star}).  The dashed lines denote
  the phase-space expected to encompass 95\% of all normal star-forming
  galaxies.  We find elevated levels of Balmer 
  absorption in the 24$\mu$m-bright cluster and group members relative to
  the field, indicating a greater incidence of starburst activity. \label{fig-EW}}
\end{figure}

For the higher S/N composite spectra, fitting techniques were used to
determine the EW of the [OII] and H$\delta$ features. The [OII] doublet was
fit using a double Gaussian model at fixed wavelength separation (2.8 \AA)
plus a linear continuum. The H$\delta$ line was also fit with a two Gaussian
model plus a linear continuum, the two Gaussians characterizing the
absorption and emission component of the feature. 

In Figure \ref{fig-EW} we plot the EW of [OII] against that of H$\delta$ measured
in the composite spectra of the 24$\mu$m-detected cluster, group and
superfield galaxies.  Also shown is the parameter space that Dressler et al.~(2009)
calculate should encompass 95\% of all normal star-forming galaxies based on
the mean relationship between EW([OII]) and EW(H$\delta$) observed in the
Sloan Digital Sky Survey and from the analysis of Goto et
al.~(2003)\footnote{Since the EW(H$\delta$) measurements used by Dressler et
  al.~(2009) for this calculation were not infill corrected, our values shown
  in Figure \ref{fig-EW} are similarly not corrected for emission infill.}.
We find that the detected superfield galaxies have on average moderate [OII] and
H$\delta$ line strengths, placing them well within the region where we expect
to find galaxies undergoing continuous star formation.  This is not the case
for the 24$\mu$m-detected group galaxies, which show an excess of Balmer
absorption indicating a greater incidence of bursty activity.  The detected
cluster members exhibit a similar excess, but only among the more
luminous galaxies which have at least LIRG-level luminosities ($L_{\rm
  TIR}>10^{11}$ $L_{\odot}$).  
When considering all detected cluster members (with all IR luminosities), the average
EW(H$\delta$) is elevated compared to the superfield sample, but consistent
with normal star formation.  When we average only the more luminous members,
we find a substantial increase in EW(H$\delta$) that is not seen in the field
population with equivalent luminosities.  

The different spectral properties we find as a function of IR luminosity
suggests the lower luminosity cluster members are undergoing high levels of
continuous star formation, while the cluster LIRGs are experiencing
more bursty activity.  
The reason we do not find a substantial change in the EW(H$\delta$) in the
group sample when we cut on IR luminosity is because almost all of the detected group
galaxies (7/8) are LIRGs and show strong Balmer absorption.

It is important to note that we do not find a similar increase in Balmer
absorption among the LIRGs in the field.  
Both the superfield LIRG sample and the entire superfield sample have average spectral
properties consistent with normal, continuous star formation.
This is largely due to H$\delta$ emission infill that is absent in the
spectra of galaxies in the denser environments.  
This indicates the cluster and group galaxies are either affected by
more dust obscuration, experiencing burstier activity, or both, as the two are
known to be correlated (Dopita et al.~2002; Kewley et al.~2004).

Although a Balmer excess can be produced through the rapid truncation of
normal star formation, all of the galaxies considered here have high IR
luminosities and inferred SFRs.  
It is more likely that this excess is the result of a temporary increase in
activity experienced within the cluster and group environments.  This finding has
important ramifications as it indicates the star forming cluster and group
galaxies are not simply analogs of the field population that have yet to be
quenched.  A greater fraction of the galaxies in the denser environments are
experiencing a burst of star formation compared to field galaxies at the same
redshift.

\subsection{Optical vs IR SFRs}

A final noteworthy point regarding the spectra of the 24$\mu$m-detected sample is the
small fraction of galaxies that have [OII] emission lines strong enough to be
classified as e(b) galaxies. This lack of strong [OII] emission has
previously been noted for IR-luminous cluster galaxies (Marcillac et
al.~2007; Dressler et al.~2008) and is thought to be the result of selective 
dust extinction, where younger stellar populations that produce the [OII]
doublet are more heavily obscured than the older A-type stars that give rise
to Balmer absorption features (Poggianti \& Wu 2000).  Poggianti et
al.~(1999) previously proposed that due to such extinction, e(a) galaxies in
higher redshift clusters could be the dusty, starbursting progenitors of the
k+a population observed in these systems, a scenario that our observations support.  

\begin{figure}
\epsscale{1.1}
\plotone{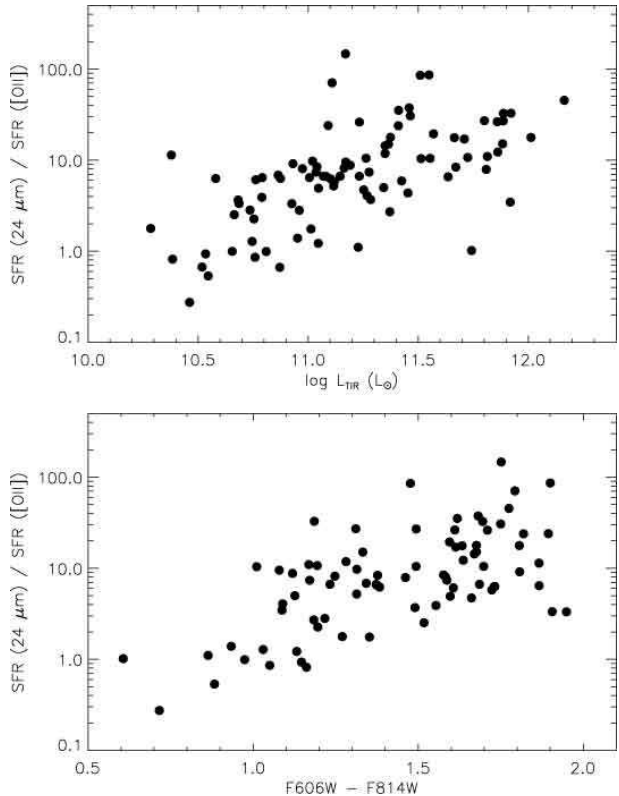}
\caption{The ratio of SFRs derived from a galaxy's 24$\mu$m luminosity and
   [OII] line luminosity for all detected supercluster members as a function
  of total IR luminosity (\emph{top panel}) and optical color (\emph{bottom
    panel}).  We find that both our most IR luminous and reddest 24$\mu$m sources 
  suffer from the greatest optical obscuration. \label{fig-SFR_comp}}
\end{figure}

We can gauge the level of dust obscuration affecting the 24$\mu$m-bright members 
by comparing their optically-derived SFRs (determined from their [OII] line
luminosities) to the more dust-insensitive rates obtained from their IR
emission.  To estimate these rates we employed the $L([\rm OII])-{\rm SFR}$ and
$L_{\rm TIR}-{\rm SFR}$ relationships of Kennicutt (1998) as described in
\S3.4 and \S4.2.  The results are shown in Figure \ref{fig-SFR_comp}, where we plot the
ratio of optical to IR-derived SFRs as a function of IR luminosity
(and optical color, see \S5.4). For any given galaxy several factors may
offset the optically-derived SFR from its IR counterpart in addition to
extinction, including variations in metallicity and ionization parameter,
which likely contribute to the significant scatter seen in Figure
\ref{fig-SFR_comp}.  Nonetheless, the overall trend is apparent: an
increasing disparity between the [OII] and 24$\mu$m-derived SFRs with increasing
IR luminosity.  This reflects the well known result that the severity of dust
extinction scales with star formation activity in IR-luminous systems (Dopita
et al.~2002; Kewley et al.~2004).  For the most IR luminous galaxies ($L_{\rm
  TIR} > 10^{11.5}$) the discrepancy between the optical and IR-derived rates is
roughly one order of magnitude.
Although the star formation activity of these galaxies is not completely
hidden by dust at optical wavelengths, the use of optical-line diagnostics
would severely underestimate the level of ongoing activity in these systems.

\section{Morphologies}

Thus far we have found an overdensity of 24$\mu$m-bright galaxies in the
Cl1604 systems and have determined that their star formation activity is
burstier in nature than their counterparts in the field.  In this section we
use our ACS imaging of these galaxies to investigate whether galaxy
interactions play a significant role in triggering their activity.
All of the 24$\mu$m-detected cluster and group members and 95\% of the
undetected members are covered by this high-resolution imaging, allowing us
to compare the morphological mix and merger fraction of the two populations.   
We classified the galaxy morphologies using both visual and automated
techniques.  The manual classification was carried out via visual inspection
of each galaxy by one of the authors (L.M.L.), while the automated
classification was carried out using the MORPHEUS
software\footnote{http://odysseus.astro.utoronto.ca/$\sim$abraham/Morpheus/},
which we used to calculate the Gini coefficient, $G$, and the $M_{20}$ parameter for
each galaxy.  The former is a measure of the symmetry in a galaxy's flux distribution (Abraham
et al.~2003), while the latter is the second-order moment of the brightest 20\% of a
galaxy's flux and a measure of central concentration (Lotz et al.~2004).   

\begin{figure}[t]
\epsscale{1.15}
\plotone{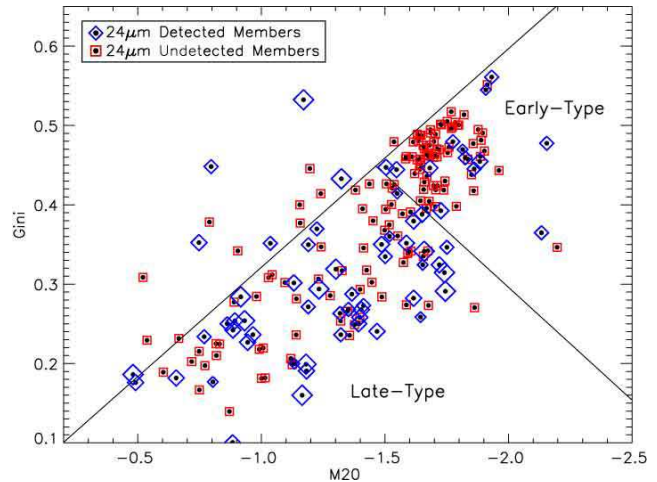}
\caption{Gini and $M_{20}$ parameters for the 24$\mu$m-detected (\emph{blue
    diamonds}) and undetected (\emph{red squares}) supercluster members.  The
  symbol size of the 24$\mu$m-bright galaxies scale with total IR
  luminosity in the same manner as shown in Figure 3. We find that the
  detected members are predominately late-type
  systems. \label{fig-gini_m20}} 
\end{figure}

Both our visual and automated classifications find that the 24$\mu$m-detected
cluster members are predominantly late-type galaxies.  This can be seen in
Figure \ref{fig-gini_m20}, which shows the $G-M_{20}$ distribution for the
24$\mu$m-detected and undetected cluster members.    Only a small fraction
of the IR-bright galaxies are found in the early-type region of the
$G-M_{20}$ parameter space and these tend to have lower IR luminosities than
their late-type counterparts ($L_{\rm TIR}$ is indicated by symbol size in
Figure \ref{fig-gini_m20}).  The opposite trend is clearly visible for the
undetected member galaxies, which largely cluster in the early-type region.   

There are several scenarios that may explain the presence of
early-type systems among the IR-bright cluster members, including 
the possibility that their 24$\mu$m emission is 
due to AGN activity missed by our SED fitting.  
That said, the supercluster AGN that were detected tend to have among the
highest IR luminosities in the cluster sample, whereas the detected early-type
galaxies have among the the lowest, making this scenario unlikely. 

\begin{figure*}[t]
\epsscale{1.15}
\plotone{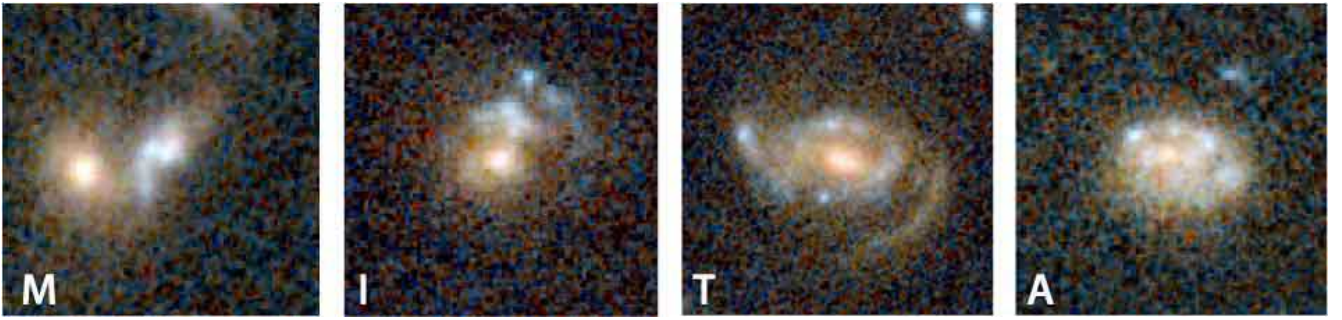}
\caption{Color ACS thumbnails showing examples of our four interaction
  classifications: M - ongoing mergers, I - galaxies that exhibit tidal
  features and have a nearby companion, T - galaxies that exhibit tidal
  features but do not have an obvious companion, and A - galaxies with
  asymmetric light distributions. The color components are F606W and F814W
  for blue and red, respectively, while green is represented by a weighted
  average of the two. \label{fig-morph_ex}} 
\end{figure*}

To better quantify the morphological mix and the incidence of merger and/or
interaction signatures among the cluster members, we have decided to rely
primarily on our visual classifications as opposed to our calculated $G$ and
$M_{20}$ parameters\footnote{This is largely because our experiments with
  MORPHEUS have found that the ability of the G and $M_{20}$ parameters to
  detect interacting galaxies depends heavily on the deblending parameters
  used to separate objects during the detection process.}.  Upon inspection,
each galaxy was assigned an early-type (E or S0), late-type (Sa through Sd)
or merger/irregular morphology, the latter of which was given to disturbed or
interacting systems that could not be assigned a standard morphology, as well
as ring galaxies.  In addition each galaxy was given an interaction
classification that represents the likelihood that the system has experienced
a recent merger or interaction event.  These classifications consist of
ongoing mergers or those likely to merge (M; typically disturbed with an
irregular companion), interaction (I; these often exhibit tidal features and
have a nearby companion), tidal disruption (T; these exhibit tidal features
and do not have an obvious companion), or asymmetric morphology (A; possibly
disturbed, these have asymmetric light distributions).  While an asymmetric
disk is by no means a clear indication of a recent interaction, we have
included this classification in case the activity of the 24$\mu$m-bright
galaxies is triggered by minor interactions that do not significantly disrupt
their morphologies.  We present our results below both with and without this
classification.  An example of each of our four interaction classifications
is shown in Figure \ref{fig-morph_ex}. 

\begin{center}
\begin{deluxetable}{llll}
\tabletypesize{\scriptsize}
\tablewidth{0pt}
\tablecaption{Morphologies of Cluster and Group Members\label{tab-morph}}
\tablecolumns{4}
\tablehead{\colhead{}  & \colhead{$R<2R_{\rm vir}$} & \colhead{$R<R_{\rm vir}$} & \colhead{$R>R_{\rm vir}$}  \\
           \colhead{Morph} & \colhead{(Det / Undet)} & \colhead{(Det / Undet)} & \colhead{(Det / Undet)}}
\startdata
E/S0        & 30.3\% / 66.7\%  & 34.2\% / 76.6\% & 25.0\% / 43.9\% \nl
Late-Type   & 62.1\% / 28.9\%  & 52.6\% / 21.3\% & 75.0\% / 46.3\% \nl
Irregular   & 07.6\% / 04.4\%  & 13.2\% / 02.2\% & 00.0\% / 09.8\%  \nl
\vspace*{-0.075in}
\enddata
\tabletypesize{\scriptsize}
\tablecomments{Det = 24$\mu$m Detected, Undet = 24$\mu$m Undetected}
\end{deluxetable}
\end{center}

\vspace*{-0.35in}
Our findings on the morphological mix of the cluster
members is summarized in Table \ref{tab-morph}, where morphological fractions
are listed for both 24$\mu$m-detected and undetected galaxies and as a function of
cluster/group-centric distance.  In general agreement with our calculated
$G-M_{20}$ distribution, we find that the IR-bright systems in the Cl1604
systems are predominantly late-type galaxies (62.1\%), while the opposite is
true of the undetected population, (67.2\% of which have E or S0 morphologies).
We find that the IR-bright population becomes slightly more bulge dominated
within the cluster/group virial radii as the late-type fraction drops to 52.6\%,
while beyond this distance the spiral fraction rises to 75\%.  A similar
trend is observed for the undetected population, whose early-type fraction
increases to 76.6\% within the virial radii and drops to 45\% in the cluster
outskirts.   

These fractions are similar when considering the cluster and group subsamples
separately, with the notable exception being the irregular morphologies. In
both the three clusters and three groups, all of the 24$\mu$m-detected
irregulars are found within the virial radius, but in the case of the groups,
the irregular fraction is equal to that of the late-type morphologies,
comprising 37.5\% of the IR-bright sample.  
If these morphologies are due to disruptive mergers, it would seem the group
environments, with their lower relative galaxy velocities, are highly
conducive to such activity.  We elaborate on this below when we discuss the
merger fraction in these systems.

In Table \ref{tab-merg} we present statistics on the interaction classifications assigned
to the cluster and group members, again for both 24$\mu$m-detected and undetected
galaxies and as a function of cluster/group-centric distance. 
The first row of Table \ref{tab-merg} presents the fraction of galaxies with either
irregular morphologies or an M classification.  The second row includes these
 as well as I and T galaxies, while the third also includes galaxies with an A
 classification.  We find that regardless of which of these three schemes we
 adopt to identify possible interactions, the 24$\mu$m-detected members are
 twice as likely to exhibit merger or interaction signatures than their
 undetected counterparts.  The fraction of detected cluster and group members
 within $2R_{\rm vir}$ with an irregular morphology or an M, I, or T classification is
 53.0\%, while only 25.9\% of the undetected members have similar
 properties\footnote{These fractions increase to 68.2\% and 32.6\%,
   respectively, if we include the A classification.}.  Interestingly, this
 difference is largely due to an increase in the fraction of disturbed
 24$\mu$m members within the cluster and group virial radii.  Outside the
 virial radii the detected and undetected populations have similar disturbed
 fractions (35.7\% versus 39.0\%, respectively),
while within the virial radius the 24$\mu$m-bright population is three times
as likely to exhibit a disturbance or merger signature (65.8\% versus
20.2\%).   We have confirmed that this is not solely due to an increased rate
of interactions in the groups, as we find similar results when we examine the
cluster and group subsamples separately. 

\begin{center}
\begin{deluxetable}{llll}
\tabletypesize{\scriptsize}
\tablewidth{0pt}
\tablecaption{Interaction classification of Cluster and Group Members\label{tab-merg}}
\tablecolumns{4}
\tablehead{\colhead{Interaction}  & \colhead{$R<2R_{\rm vir}$} & \colhead{$R<R_{\rm vir}$} & \colhead{$R>R_{\rm vir}$}  \\
           \colhead{Class.} & \colhead{(Det / Undet)} & \colhead{(Det / Undet)} & \colhead{(Det / Undet)}}
\startdata
Irr+M       & 34.9\% / 18.5\% & 42.1\% / 13.8\% & 25.0\% / 29.3\% \nl
Irr+MIT     & 53.0\% / 25.9\% & 65.8\% / 20.2\% & 35.7\% / 39.0\% \nl
Irr+MITA    & 68.2\% / 32.6\% & 78.9\% / 26.6\% & 53.6\% / 46.3\% \nl
\vspace*{-0.075in}
\enddata
\tabletypesize{\scriptsize}
\tablecomments{Det = 24$\mu$m Detected, Undet = 24$\mu$m Undetected}
\end{deluxetable}
\end{center}

\begin{figure*}
\epsscale{1.1}
\plotone{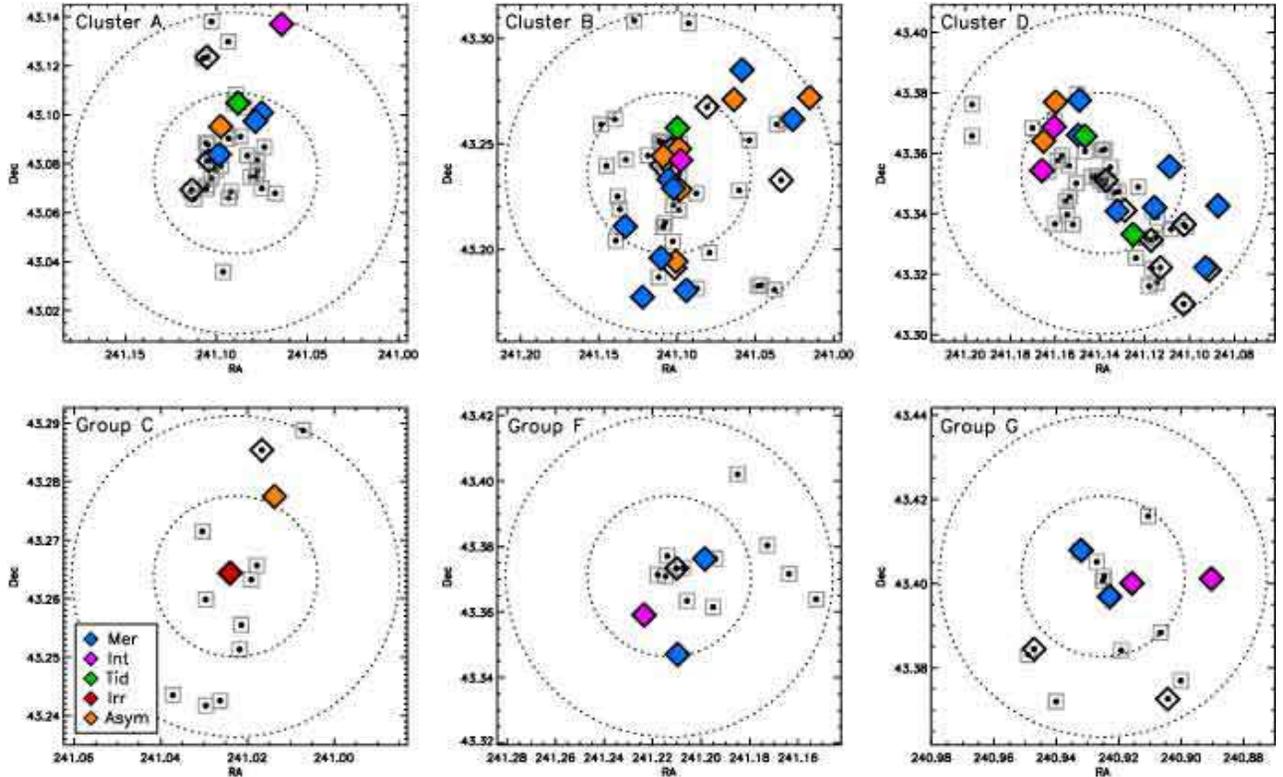}
\caption{Spatial distribution of 24$\mu$m-detected (\emph{blue diamonds}) and
  undetected (\emph{red squares}) cluster and group members.  Each
  24$\mu$m-detected galaxy with a disturbed morphology is color coded
  according to its morphology or interaction classification as indicated by
  the legend.  \label{fig-radec_morph}}
\end{figure*}

\vspace*{-0.35in}
A graphical representation of our morphological analysis can be seen in Figure
\ref{fig-radec_morph}, which shows the spatial distribution of the disturbed
24$\mu$m population within each cluster and group.  In this figure each
galaxy has been color coded based on its morphology and/or interaction
classification.  This plot highlights three notable aspects of our findings.
First, we find that nearly all (7/8) of the detected group members within $R_{\rm
  vir}$ exhibit merger or interaction features.  Keeping in mind the low
number statistics, this suggests that among the group members that are
experiencing high levels of current star formation, a vast majority have
experienced a merger or interaction in the recent past.  This agrees with our
earlier assessment, based on the fraction of irregular morphologies, that the
groups systems appear conducive to such activity.

Second, although we find several galaxies with merger signatures in the
filament connected to Cluster D, many other galaxies in this region exhibit no such features.
This includes galaxies with some of the highest SFRs detected in the supercluster.
Therefore we must conclude that the high level of activity seen in the
filament is not solely a result of disruptive galaxy interactions as systems infall toward the
cluster environment.

Third, there are indications that many of the disturbed 24$\mu$m-bright cluster members
have interacted with the core of their host clusters.  For example, in
Cluster D, all of the 24$\mu$m detected members to the northeast of the
cluster center show some form of morphological disturbance.  Given the alignment of these
galaxies with the infall direction of the filament, we have interpreted these
systems to be galaxies that have already passed through the cluster core.
Likewise in Cluster B, a substantial fraction of the 24$\mu$m
detected galaxies located at low cluster-centric distances exhibit
disturbed morphologies.  
This may be due to increased tidal interactions as galaxies move through the
cluster centers.

In summary, we find that i.) the 24$\mu$m-detected supercluster members are
primarily late-type galaxies, however in the group environment irregular
morphologies are as prevalent as spirals, ii.) the fraction of
24$\mu$m-detected cluster and group members that exhibit disturbed morphologies
or merger signatures is 65.8\% within $R_{\rm vir}$ and 35.7\% outside
$R_{\rm vir}$.  The former is three times the fraction observed in the
undetected population, while the latter is roughly equal to it.
iii.) Most detected galaxies near the cluster cores have disturbed
morphologies, while many in the filament extending from Cluster D show no
such disturbances.

\section{Optical Colors}

\begin{figure}
\epsscale{1.15}
\plotone{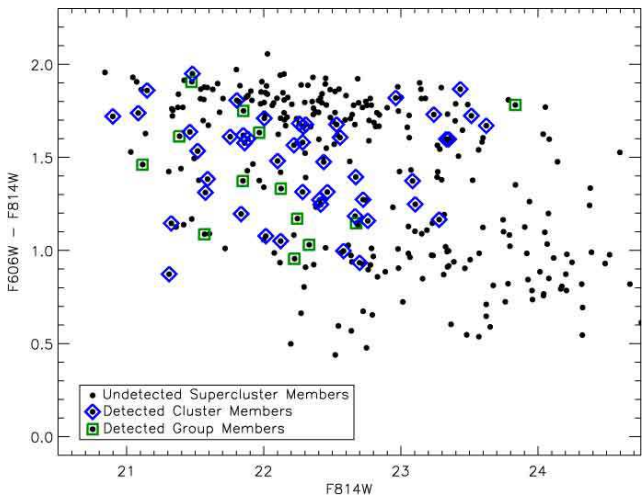}
\caption{Optical color (F606W-F814W) as a function of apparent magnitude in
  the F814W band for all supercluster members (\emph{black points}).
  $24\mu$m-detected cluster members are indicated by blue diamonds, while
  detected group members are denoted by green squares.\label{fig-color-mag}}
\end{figure}

In addition to our morphological analysis, we have used our dual-band ACS
imaging of the Cl1604 systems to examine the integrated optical colors of the
IR-bright cluster population.  A color-magnitude diagram containing both
24$\mu$m-detected and undetected cluster and group members is shown in
Figure \ref{fig-color-mag} in the F606W and F814W bands. At the
cluster redshifts, these ACS filters sample the rest-frame U and B bands,
respectively. In general we find that the 24$\mu$m-detected galaxies tend to
be redder than their undetected counterparts in the blue cloud and yet
slightly bluer than the supercluster red sequence.  
This color distribution, peaking in the so-called green
valley, has been previously observed for 24$\mu$m-selected galaxies in both
cluster and field environments (Geach et al.~2006; Cowie et al.~2008; Krick
et al.~2009) and has been shown to be largely the result of severe dust extinction in
otherwise blue, star forming systems.  We discuss the effects of dust on the
colors of these galaxies in greater detail at the end of this section.

\begin{figure}
\epsscale{1.1}
\plotone{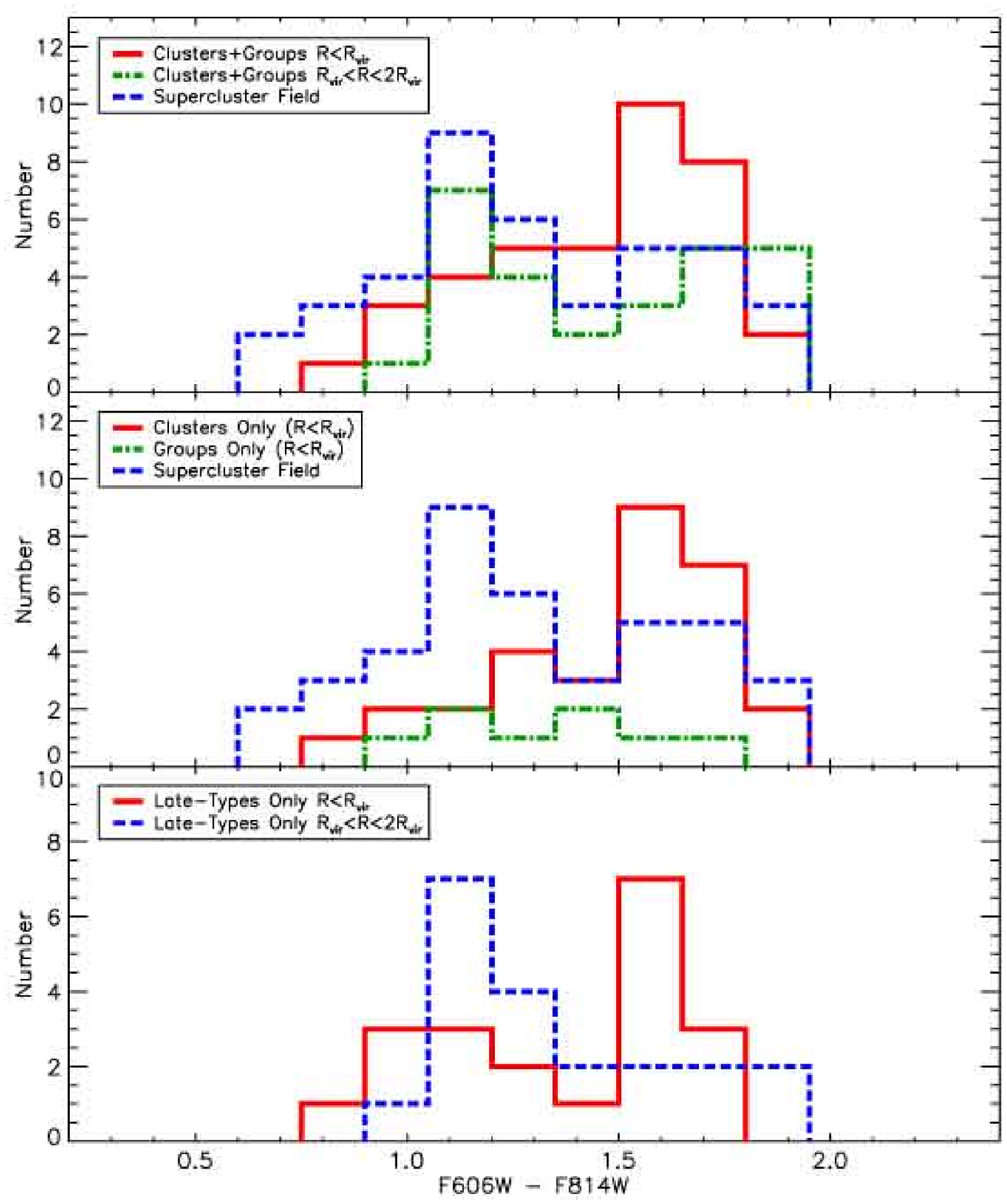}
\caption{Optical color (F606W-F814W) distribution of $24\mu$m-detected
  cluster, group and superfield galaxies.  The top panel shows the color
  distribution as a function of cluster- and group-centric distance.  In the
  middle panel the distribution is shown for cluster and group galaxies
  within $R_{\rm vir}$ separately. The bottom panel shows the colors of
  detected late-type cluster members as a function of cluster-centric distance.
  In general we find redder colors for detected cluster members at
  lower cluster-center distances compared to galaxies on the cluster
  outskirts and those in the superfield.
  This is true even when considering only galaxies confirmed
  to have late-type morphologies. \label{fig-color-hist}}
\end{figure}

Interestingly, when we split the 24$\mu$m-detected sample by cluster/group-centric
distance we find a pronounced segregation in optical color.  Figure
\ref{fig-color-hist} shows the color distribution of 24$\mu$m-bright cluster
and group galaxies located within the virial radius ($R<R_{\rm vir}$), on the
cluster and group outskirts ($R_{\rm vir} < R < 2R_{\rm vir}$), and galaxies
in the supercluster field ($R>2R_{\rm vir}$) .  We find that in the field, the
24$\mu$m-detected galaxies have colors that peak near the blue cloud, with a
smaller subset exhibiting relatively red colors.  On the cluster and group
outskirts ($R_{\rm vir} < R < 2R_{\rm vir}$) the color distribution is fairly
even, but within $R_{\rm vir}$ the distribution has reversed and the 24$\mu$m
galaxies become predominantly redder, with a smaller subset of blue systems.
As the middle panel of Figure \ref{fig-color-hist} shows, this blue tail is
largely due to galaxies in the Cl1604 groups.  When we split the $R<R_{\rm
  vir}$ sample into group and cluster subsamples, we find a pronounced
dichotomy between the cluster and field galaxy colors.

We have confirmed that this color difference is not due to an increase in the
early-type fraction of 24$\mu$m sources within $R_{\rm vir}$, as the
disparity remains when we consider only galaxies securely identified as Sa-Sd by visual
inspection.  The color difference between confirmed late-type cluster and
group members within $R_{\rm vir}$ and those outside this distance can be seen
in the bottom panel of Figure \ref{fig-color-hist}. 
This difference is significant as it suggests the 24$\mu$m-detected galaxies
found near the cluster centers cannot simply be infalling, IR-bright field
galaxies that have yet to be quenched.   Either the cluster members are
affected by significantly more dust extinction or these galaxies have already
experienced a degree of color transformation.  

\begin{figure*}
\epsscale{1.15}
\plotone{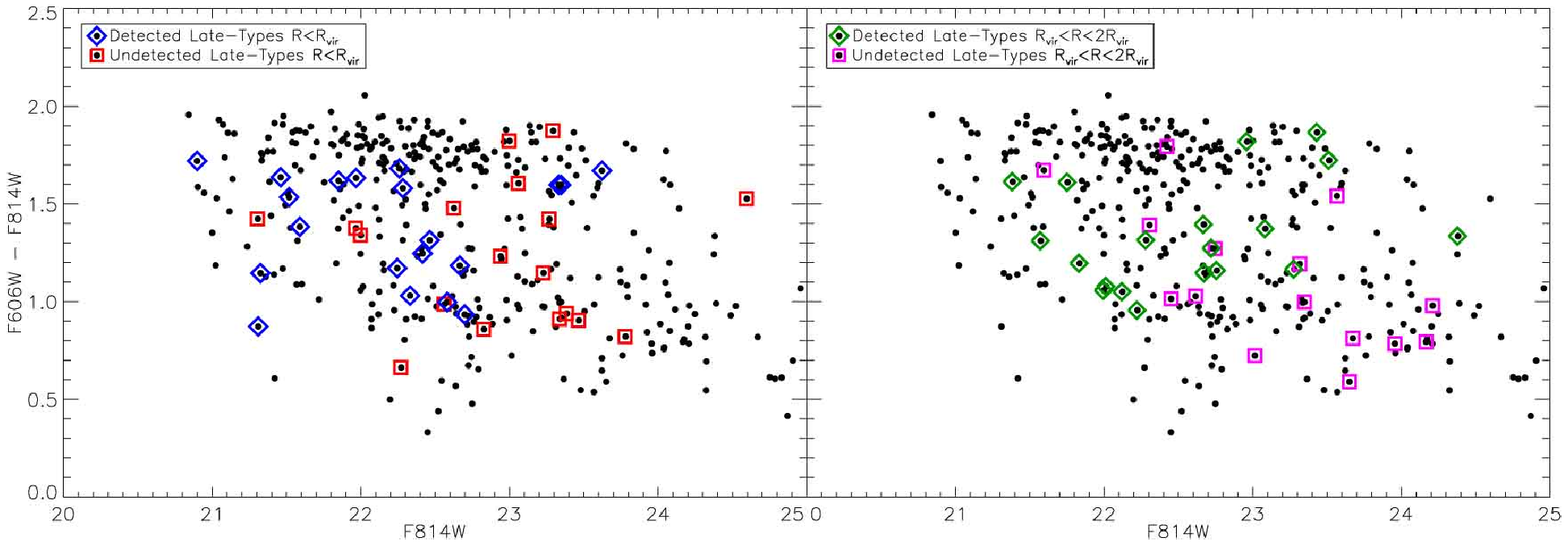}
\caption{Optical color (F606W-F814W) as a function of apparent magnitude in
  the F814W band for all supercluster members (\emph{black points}).  The
  left panel shows the colors of detected (\emph{blue diamonds}) and
  undetected (\emph{red squares}) late-type galaxies in the cluster
  cores.  The right panel is the same but for detected (\emph{green diamonds}) and
  undetected (\emph{magenta squares}) late-type
  galaxies in the cluster outskirts. \label{fig-color-mag-spirals}}
\end{figure*}

Redder colors for late-type cluster members relative to the field were
previously observed by Homeier et al.~(2006) in several high-redshift systems
(including two of the Cl1604 clusters).  They also conclude that such cluster spirals 
could not be a pristine infall population due to this color difference.
However this has been challenged by Marcillac et al.~(2007) who note that the
optical colors of 24$\mu$m-detected, late-type galaxies in RX J0152.7-1357
appear to be unaffected by their starburst activity as they have the same
colors as undetected spirals in the system.   Contrary to their result, we
find that the undetected late-type members in the Cl1604 systems are
substantially bluer and fainter than the 24$\mu$m-detected members.  This can
be seen in Figure \ref{fig-color-mag-spirals}, which shows the
color-magnitude distribution of both samples.  Whether it be within the
virial radius or on the cluster outskirts, the detected members tend to be
redder than their undetected spiral counterparts. 

It is well known that starburst activity is associated with significant
levels of dust extinction which could explain the color differences observed
for the detected and undetected populations.  For example, Cowie et
al.~(2008) demonstrate that when corrected for extinction, the bulk of the
24$\mu$m sources in the GOODS-North field out to $z\sim1.5$ move from the green
valley back onto the blue cloud.
We explored this possibility by comparing the optical and IR-derived SFRs of
the 24$\mu$m population as a function of color in the bottom panel
of Figure \ref{fig-SFR_comp}. We find a clear correlation between optical
obscuration and observed color, with the reddest sources consistently showing
the largest offset between the two SFR measures.  The severe obscuration indicated in Figure
\ref{fig-SFR_comp} could easily explain the color difference found in the two
populations.  
This result suggests that the 24$\mu$m-detected cluster members are
experiencing dustier star formation activity relative to the field population
with similar IR luminosities.  This result is consistent with our spectroscopic
analysis which found burstier activity occurring in the cluster and group
members.   These findings strongly suggest the starburst population within the clusters
and groups are not simply infalling field galaxies as they are  
experiencing a markedly different type of star formation activity.

\begin{figure}
\epsscale{1.15}
\plotone{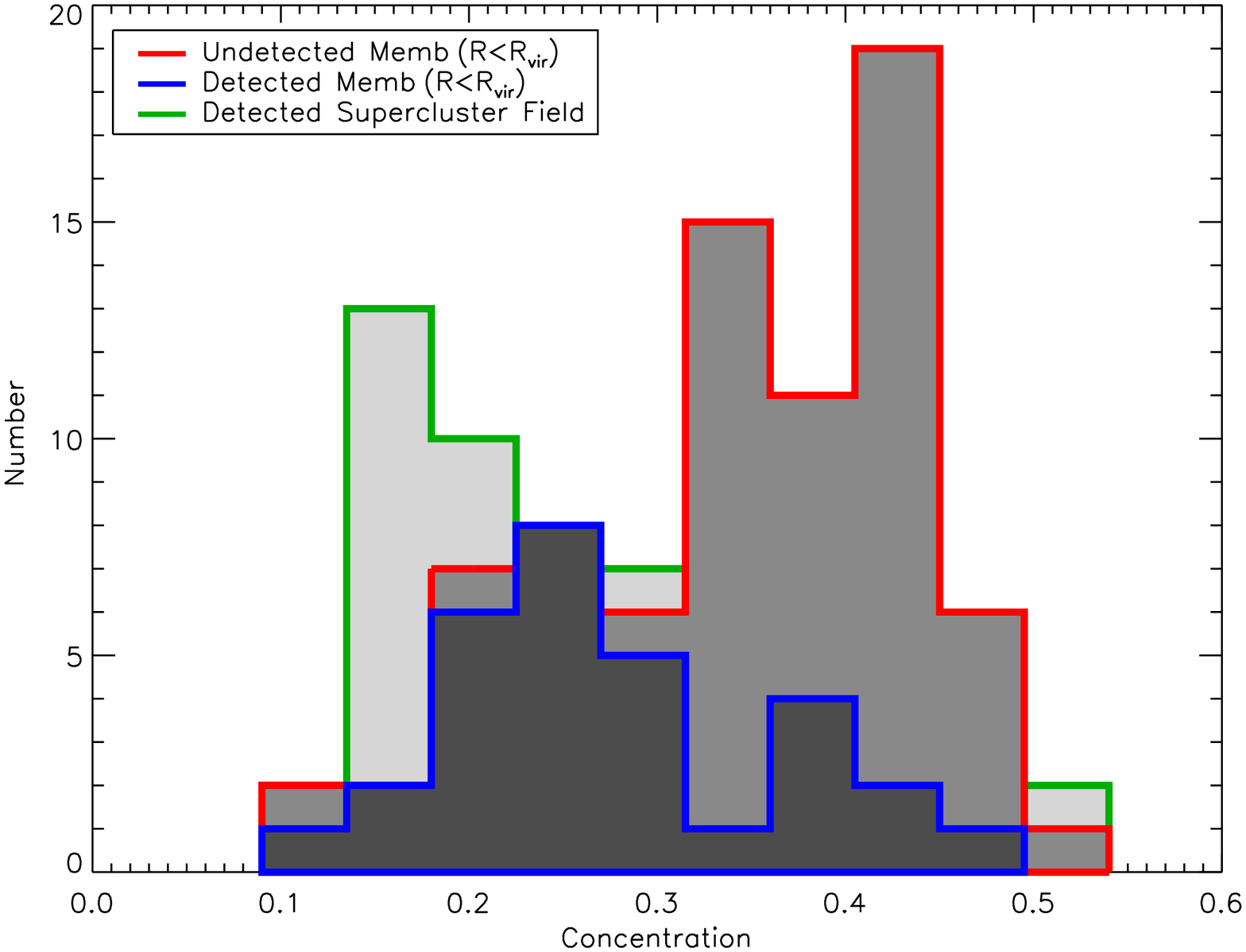}
\caption{Distribution of concentration parameters as measured by MORPHEUS for
  detected and undetected cluster and group galaxies within $R_{\rm vir}$ and
  for galaxies in the surrounding field sample.  We find the detected cluster and
  group galaxies have more centrally concentrated light distributions than
  their counterparts in the field, but not as concentrated as the undetected
  members, which are largely early-type systems.
  \label{fig-color-conc-hist}}
\end{figure}

The redder colors and increased extinction of the IR-bright members within
the virial radius could be explained if they were experiencing more centrally
concentrated bursts of star formation, which tend to be buried beneath higher
column densities of obscuring dust.  Evidence for such activity has been
observed in other IR-bright cluster galaxies.  For example, Geach et
al.~(2009), using the \emph{Spitzer} Infrared Spectrograph, find that the
integrated spectra of LIRGs in Cl0024+16 resemble those of local nuclear starbursts.
If such activity is responsible for the redder colors we observe among the
Cl1604 galaxies, we may expect to find a more centrally concentrated
distribution of light in these galaxies.    
To test this, we measured the concentration parameter for each 24$\mu$m-detected
galaxy using the MORPHEUS software and indeed we find a higher concentration
for galaxies within $R_{\rm vir}$ versus those in the supercluster field, as
well as for the redder 24$\mu$m-detected population in general.  This can be
seen in Figure \ref{fig-color-conc-hist}, which shows the measured
concentration parameter for various subsamples within the Cl1604 system.
While all of the detected galaxies in general have low concentrations, we find the
cluster/group distribution peaks at higher concentrations than that of the
field galaxies.

Although there is a correlation between concentration and Sersic index, we have
confirmed that the higher concentration parameters found for the redder
populations are not simply due to more early-type galaxies associated with the
24$\mu$m-sources within $R_{\rm vir}$.  While the IR-bright members
within $R_{\rm vir}$ have higher concentrations than those in the field,
their concentrations fall far below that of the undetected, early-type
cluster and group members, indicating that 
this result is not due to a preponderance of early-type morphologies.

In summary, we find that i.) the 24$\mu$m-detected cluster members within $R_{\rm
  vir}$ have considerably redder optical colors than their IR-bright
counterparts in the field, ii.) the reddest of these sources exhibit the
largest offset between their optically- and IR-derived SFRs, indicating a
greater level of optical obscuration, and iii.) the 24$\mu$m-detected cluster
and group members have higher central concentration of rest-frame B-band light compared
to detected field galaxies.

\section{Discussion}

\subsection{Infalling Field Galaxies or Triggered Activity?}

One of the central issues debated since the initial detection of the
Butcher-Oemler effect is whether galaxies experience a period of enhanced
star formation during cluster assembly.  Despite spectroscopic evidence for
starburst activity in distant clusters, the answer to this question has
remained elusive since the evolution of the star forming properties of
cluster galaxies often mimics the evolution found in the field (e.g.~Bai et al.~2007).  
In our study of the Cl1604 clusters we have found evidence that the elevated
star formation activity in these systems is not simply a reflection of the
increased field activity at $z\sim0.9$.  In summary, we find an overdensity
of IR-bright galaxies in the Cl1604 clusters relative to the coeval field and
that this excess scales with a system's dynamical state.  These galaxies
exhibit elevated levels of Balmer absorption and increased levels of
obscuration which suggests their star formation activity is burstier than
that of field galaxies with similar IR luminosities.   

Our findings regarding the excess of 24$\mu$m-bright galaxies in dynamically
active systems is consistent with several previous studies that have reported
source overdensities in unrelaxed or merging clusters (Geach et al.~2006, Marcillac et
al.~2007).  Recently Saintonge, Tran, \& Holden (2008), using MIPS observations of eight
clusters, found that while the fraction of galaxies undergoing dust-obscured
star formation steadily increases with cluster redshift, dynamically
unrelaxed systems exhibit an excess of sources beyond the nominal redshift
trend.  The authors attribute this increase to a higher galaxy accretion rate in
unrelaxed clusters.  Our results generally agree with this assessment given that the
24$\mu$m detected galaxies appear to be an infalling population. 
This is especially apparent in Cluster D, where we find dust-obscured
activity associated with galaxies streaming along a large-scale filament
connected to the system.

That being said, the differences we observe in the IR-bright cluster
members relative to the field indicate that their star formation activity
cannot simply be residual field activity that has yet to be quenched.
Their optical colors are generally redder (which we have shown correlates
well with extinction in this sample), their morphologies are more
centrally concentrated, and their spectra show signs of elevated starburst
activity relative to the field.   Dressler et al.~(2004) previously
demonstrated that the average spectral properties of moderate-redshift
clusters could not be accounted for through a mix of passive and
continuously star-forming galaxies and that starburst activity must play a
larger role at higher redshifts.  However, they found this to be true of the
field population as well.  In our study, we find that the composite spectra
of the cluster and group members show a greater contribution from starburst galaxies
than we see in the field population.  When one further considers that the
lifetime of a burst ($\sim500$ Myr) is far shorter than the time required for
an infalling galaxy to reach the cluster core (1260 Myr in cluster D for a
galaxy infalling from the virial radius at 582 km s$^{-1}$), it becomes evident that
the elevated starburst activity we observe must be triggered, to some
extent, by processes related to the cluster environment.
This finding is at odds with scenarios that simply rely on an increased
rate of passive accretion to explain the greater incidence of star forming
galaxies in higher redshift clusters (e.g.~Ellingson et al.~2001, Loh et al.~2008).

\subsection{Triggering Mechanisms}

We have presented three observational clues that can help shed light on the
mechanism responsible for initiating the elevated starburst activity we observe:
i.) the fraction of 24$\mu$m-detected galaxies showing signs of a recent 
merger or tidal interaction is nearly twice as high within the cluster/group
virial radius (65.8\%) than it is outside this distance (35.7\%), ii.)
interaction signatures are three times as common among the detected members
than the undetected members, iii.) the 24$\mu$m-detected sample within the
virial radius exhibit a higher central concentration of rest-frame B-band
light than coeval field galaxies with equivalent IR luminosities. 

Galaxy mergers and weak tidal interactions (i.e. harassment) are two
mechanisms that can both produce the disturbed morphologies we observe and
trigger episodes of enhanced star formation (Barnes \& Hernquist 1991, Moore
et al.~1996; Barton et al.~2007).  Given the increased frequency of disturbed
morphologies among the 24$\mu$m-detected members, it is likely interactions
play a role in triggering the increased activity we observe.  This cannot be
said of the IR-luminous field population at similar redshifts.  While
LIRG-level activity at low redshifts is predominantly associated with galaxy
interactions (Ishida et al.~2004), this is not the case at the redshift of
the Cl1604 systems.  For example, Melbourne et al.~(2005) report that a
majority of LIRGs in the field at $z\sim1$ appear to be undisturbed late-type
systems rather than galaxies with peculiar or irregular morphologies.  
We also observe this for galaxies on the cluster/group outskirts and in the
supercluster field, but find the opposite result within the cluster/group
centers, where a majority of the 24$\mu$m-detected galaxies exhibit
morphological disturbances.  If continuous, LIRG-level activity is the normal
mode of star formation for gas-rich field galaxies at $z\sim0.9$ and
triggered starburst activity more prevalent in denser environments,
then this could explain many of the physical differences we observe in our
24$\mu$m-detected sample as a function of cluster/group-centric distance. 

Can interactions explain our third point: higher central concentrations for
the 24$\mu$m-detected galaxies in the cluster and group centers? 
Simulations have shown that both numerous tidal interactions and mergers can funnel
material to the center of a galaxy and trigger a circumnuclear starburst
(e.g.~Noguchi 1988; Hernquist 1989; Mihos \& Hernquist 1996). 
Observationally, this effect has been detected in several low redshift clusters.
In a survey of H$\alpha$ emission from late-type galaxies in eight
Abell clusters, Moss \& Whittle (2000) found that circumnuclear
starburst activity is preferentially found in regions of high galaxy
densities and associated with galaxies that have peculiar or distorted
morphologies.  They conclude that tidal interactions, working more
efficiently near the cluster centers, are responsible for both the disturbed
morphologies and the compact star formation.
This is in excellent agreement with our findings as we observe both an increase
in starburst activity and interaction signatures at low cluster-centric radii.  

Since both mergers and harassment can give rise to many of the morphological
disturbances we find, our observations cannot rule out either mechanism as
the starburst catalyst. Although major mergers are not expected to be common
within the cluster environment given the high relative velocities of cluster
galaxies, Oemler et al.~(2009) recently found evidence for merger activity
associated with 24$\mu$m-detected galaxies near the core of Abell 851.
Citing Struck (2006), they proposed that the compression of recently accreted
groups and bound galaxy pairs could lead to an increased merger rate and
elevated SFRs among infalling galaxies.
In the Cl1604 clusters we do observe signatures of galaxy-galaxy mergers that
could not originate from harassment alone, such as interacting pairs and ring
galaxies, indicating that merger activity is present to some extent in these
high density environments.  That being said, the increase in disturbed
morphologies we find within the cluster centers is largely driven by an
increase in galaxies exhibiting tidal features and not those involved in
obvious mergers.  The fraction of 24$\mu$m-detected cluster members that have
either an irregular morphology or M interaction classification is the same
within the cluster cores as it is on the cluster outskirts\footnote{This is in
  contrast to the Cl1604 groups, where such morphologies are found in
  excess.}.  On the other hand, the fraction of detected galaxies showing
tidal features (both with and without a nearby companion) is substantially
higher within the cluster centers. 
This is not to say harassment is the dominate mechanism affecting these
galaxies since bone-fide mergers may also be responsible for generating many
of the tidal features we observe.  Considering further that mergers are not
always obvious in optical imaging, it is difficult to judge the relative
importance of the two mechanisms other than to say signatures of both are
common among our 24$\mu$m-detected cluster sample. 

In the case of the Cl1604 group sample, the picture is a bit clearer.  Among
the IR-bright group galaxies we find that irregular morphologies are as
common as late-type morphologies and that almost all detected galaxies within
$R_{\rm vir}$ (7/8) show clear signs of ongoing mergers or interactions.
These interactions appear to be having a substantial effect on the star
formation activity of the detected group galaxies, as they have among the
highest IR luminosities in the supercluster sample and exhibit a large
Balmer excess relative to both IR-bright field galaxies and the detected
cluster population.  Therefore, while a mix of harassment and mergers are
likely driving the enhanced star formation found in the Cl1604 clusters,
mergers appear to be the dominate triggering mechanism in the group
environment, where the lower relative velocity of galaxies is much more
conducive to such activity.

\section{Conclusions}

We have used \emph{Spitzer}-MIPS $24\mu$m imaging, \emph{Keck}
spectroscopy and \emph{HST}-ACS observations to study the obscured star forming
population in and around three clusters and three groups at $z\sim0.9$.
These six systems are components of the Cl1604 supercluster, the largest
structure imaged by \emph{Spitzer} at redshifts approaching unity. Our
analysis has found that the average density of 24um-detected galaxies within
the virial radius of the three Cl1604 clusters is nearly twice that of the
surrounding field and that this overdensity scales with the cluster's
dynamical state. These galaxies often appear optically unremarkable and
exhibit only moderate [OII] line emission due to severe obscuration.  
Their spatial distribution suggests they are an infalling population, but an
examination of their spectral properties, morphologies and optical colors
indicate they are not simply analogs of the field population that have yet to
be quenched.  Using stacked composite spectra, we find the $24\mu$m-detected
cluster and group galaxies exhibit elevated levels of Balmer absorption
compared to what is expected from normal, continuous star formation activity.
A similar excess is not observed in field galaxies with equivalent IR
luminosities.  In addition, these galaxies have redder optical colors and a
more centrally concentrated light distribution. 

Our interpretation of these findings is that a greater fraction of the
detected cluster and group members are experiencing nuclear starburst
activity compared to their counterparts in the field. 
Since the cluster members appear to be an infalling population, this would
suggest gas-rich galaxies at high redshift experience a temporary increase in their star
formation activity as they assemble into denser environments. 

Using \emph{HST}-ACS imaging we find that disturbed morphologies are common
among the 24$\mu$m-detected cluster and group members and become more prevalent in
regions of higher galaxy density. Detected members within the virial radius
of these systems show signs of recent interactions three times more often
than undetected members and twice as often as detected galaxies on the
cluster/group outskirts.  This is true for both the cluster and group
subsamples.  
Nearly all of the detected group members show signs of clear merger activity,
while an excess of tidal features is observed  among the cluster galaxies.
We conclude that mergers are the dominate triggering mechanism responsible
for the enhanced star formation found in the Cl1604 groups, while a mix of
harassment and mergers are likely driving the activity of the cluster
galaxies, although we cannot conclusively determine the relative importance
of the two mechanisms.  

We are continuing our spectroscopic observations of the Cl1604 systems with
the goal of compiling a highly complete redshift sample for galaxies in and
around the dynamically unrelaxed Cluster D.  Using this sample, we plan to
study both the obscured star formation - density relationship in this system
and whether the descendent of the cluster's 24$\mu$m-bright population are
the bulge-dominated lenticular galaxies found in local clusters.  We expect
to complete both studies in the near future.

\clearpage

\citestyle{aa}
\bibliography{ms.bbl}

\end{document}